\documentclass[english,prb,twocolumn,superscriptaddress,showpacs]{revtex4}
\usepackage[T1]{fontenc}
\usepackage[latin1]{inputenc}
\usepackage{verbatim}
\usepackage{amsmath}
\usepackage{graphicx}
\usepackage{amssymb}

\makeatletter

\makeatletter

\usepackage{bm}

\renewcommand{\mathbf}{\bm}


\makeatother

\usepackage{babel}
\makeatother

\begin{document}

\title{Dissipation-driven phase transitions in superconducting wires}

\author{Alejandro M. Lobos}

\affiliation{DPMC-MaNEP, University of Geneva, 24 Quai Ernest-Ansermet CH-1211
Geneva, Switzerland.}

\author{An{\'i}bal Iucci}

\affiliation{Instituto de F{\'i}sica de La Plata (IFLP) - CONICET and Departamento
de F{\'i}sica, Universidad Nacional de La Plata, cc 67, 1900 La Plata,
Argentina.}

\affiliation{DPMC-MaNEP, University of Geneva, 24 Quai Ernest-Ansermet CH-1211
Geneva, Switzerland.}

\author{Markus M{\"u}ller}

\affiliation{The Abdus Salam International Centre for Theoretical Physics, P.O.
Box 586, 34100 Trieste, Italy}

\author{Thierry Giamarchi}

\affiliation{DPMC-MaNEP, University of Geneva, 24 Quai Ernest-Ansermet CH-1211
Geneva, Switzerland.}

\date{11/16/2009}

\begin{abstract}
We report on the reinforcement of superconductivity in a system consisting
of a narrow superconducting wire weakly coupled to a diffusive metallic
film. We analyze the effective phase-only action of the system by
a perturbative renormalization-group and a self-consistent variational
approach to obtain the critical points and phases at $T=0$. We predict
a quantum phase transition towards a superconducting phase with long-range
order as a function of the wire stiffness and coupling to the metal.
We discuss implications for the DC resistivity of the wire. 
\end{abstract}



\maketitle

\section{\label{sec:intro}Introduction}

The interplay between fluctuation and dissipation phenomena in quantum
systems is presently under intensive research. Fluctuations are particularly
strong in low dimensions, as reflected by the lack of long-range order
in 1D-systems with short-range interactions \citep{mermin_wagner_theorem,giamarchi_book_1d}.
On the other hand, dissipation counteracts fluctuations effects, decreasing
the lower critical dimension \citep{caldeira_leggett, bray82_macroscopic_quantum_tunneling,chakravarty82_macroscopic_quantum_tunneling, schmid83_diffusion_dissipative_quantum_system}.

Some physical realizations of dissipative low-dimensional systems
are the well-known resistively shunted Josephson junctions arrays,
where the effect of local ohmic dissipation has been intensively studied
\citep{schoen90_review_ultrasmall_tunnel_junctions,fazio01_review_superconducting_networks,  bobbert92_transitions_dissipative_josephson_chains,goswami06_josephson_array,refael07_SN_transition_in_grains&nanowires},
superconducting grains embedded in metallic films \citep{feigelman98_smt_in_2d_proximity_array,feigelman00_keldysh_action_superconductor, spivak01_quantum_superconductor_metal_transition, feigelman01_smt_proximity_array}
and Luttinger liquids coupled to dissipative baths \citep{castroneto97_open_luttinger_liquids,cazalilla06_dissipative_transition,artemenko07_longrange_order_in_1d}. 

Narrow superconducting wires with diameter $d\ll\xi_{0}$ (where $\xi_{0}$
is the bulk superconducting coherence length) are low-dimensional
systems in which strong fluctuations of the order parameter affect
low-temperature properties.

It was originally suggested by Little \citep{little67_phase_slip},
and subsequently discussed by Langer and Ambegaokar (LA)\citep{langer67}
and McCumber and Halperin (MH)\citep{mccumber70} , that resistivity
$\varrho\left(T\right)$ in thin wires would be finite for all temperatures
below the bulk critical temperature $T_{c}$. Thermal fluctuations
cause the magnitude of the order parameter to temporarily vanish at
some point along the wire, allowing its phase to slip by $2\pi$ (the
so-called thermally activated phase-slips) and dissipate through the
Josephson relation $V=\hbar/2e\;\Delta\dot{\theta}$, where $\Delta\theta$
is the phase difference across the wire.

According to the LA-MH theory, thermal fluctuations induce a resistivity
$\varrho\left(T\right)\sim\Omega\left(T\right)\exp\left[-\Delta F_{0}/T\right]$,
where $\Delta F_{0}$ is the Ginzburg-Landau free-energy barrier between
different current-carrying states in the wire, and $\Omega\left(T\right)$
is an algebraically decreasing function of $T$. However, deviations
from the LA-MH theory in the regime $T\ll T_{c}$ were first observed
by Giordano \citep{giordano88_evidence_of_qps, giordano94} and more
recently by other experimental groups \citep{lau01, tian05_dissipation_in_Sn_wires, zgirski05_breakdown_of_superconductivity_in_wires, altomare06_experimental_evidence_of_qps, bollinger08_sit_diagram_for_wires},
leading to the conclusion that for very thin wires at low temperatures
current-decay was produced by macroscopic quantum tunneling of the
phase of the order parameter through the same free-energy barriers
(the so-called quantum phase-slips), leading to a much weaker dependence
of the resistivity on $T$. 

Moreover, it is believed that the destruction of the superconducting
state in very thin wires occurs through the proliferation of quantum
phase slips/anti phase slips pairs\citep{giamarchi_attract_1d,giordano94,zaikin97,bezryadin00,lau01,buchler04_sit_finite_length_wire,altomare06_experimental_evidence_of_qps,arutyunov08_superconductivity_1d_review},
in what constitutes the quantum analog in 1+1 dimensions to the classical
Berezinskii-Kosterlitz-Thouless\citep{kosterlitz_thouless} transition
in two space dimensions.

\begin{figure}[h]
\includegraphics[bb=100bp 300bp 500bp 480bp,clip,scale=0.6]{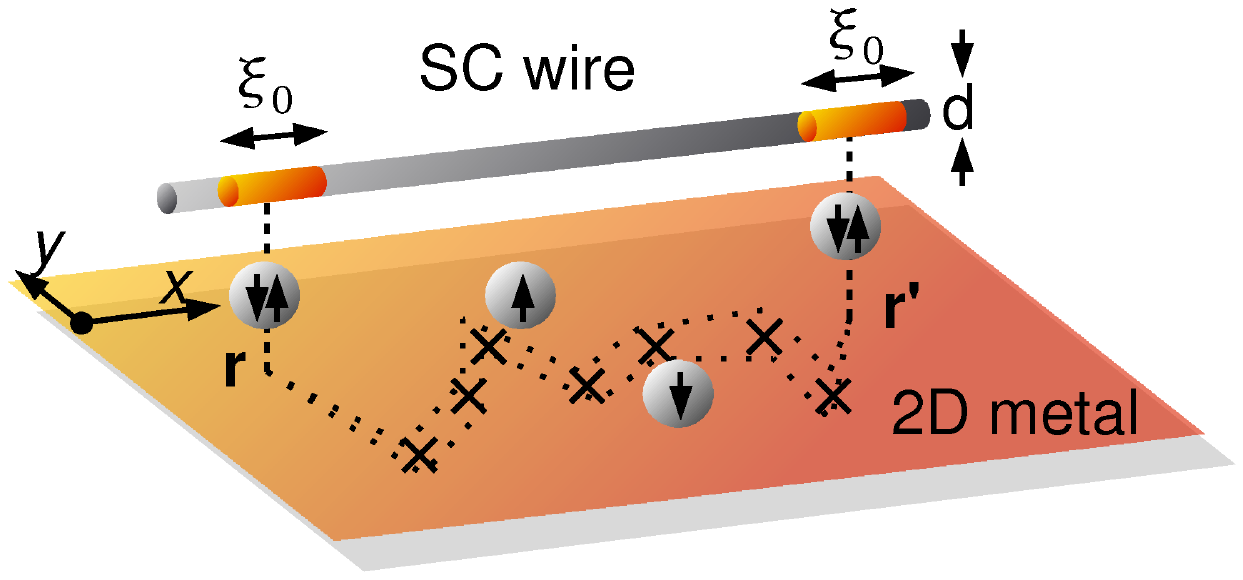}

\caption{(color online) Representation of the system. At $T\ll T_{c}$ one-particle
hopping is suppressed by the BCS gap-energy $\Delta_{0}$. At next
order in the hopping process, Cooper pairs can tunnel into the metal
and propagate coherently in the metal, generating an effective coupling
$\sim\cos\left(\theta\left(\mathbf{r}\right)-\theta\left(\mathbf{r}^{\prime}\right)\right)$
in the wire.\label{fig:schema}}

\end{figure}

Contrary to other 1D systems such as dissipative ohmic Josephson junction
arrays, isolated thin wires do not present significant sources of
dissipation at $T\ll T_{c}$ \citep{zaikin97,arutyunov08_superconductivity_1d_review}.
However, additional sources might be provided by a coupling to the
environment, a possibility which has hardly been explored yet. Although
general theoretical frameworks have been proposed to describe superconductor-normal
(SN) junctions \citep{hekking93,hekking94_andreev_conductance, feigelman00_keldysh_action_superconductor},
recent advances in superconducting nanowires fabrication techniques
call for a more detailed analysis of the effects of coupling to general
dissipation sources \citep{giordano88_evidence_of_qps, giordano94,bezryadin00,lau01,altomare06_experimental_evidence_of_qps}.

In this article we focus on the effect of weakly coupling a superconducting
wire to a diffusive 2D normal metal. We show how the induced dissipation
stabilizes superconductive long-range order at $T=0$ despite the
1D nature of the wire. At finite $T$, the effect of dissipation are
manifested in an increase of the superconductive stiffness of the
wire. 

The article is organized as follows: in Section \ref{sec:model} we
derive the effective low-energy phase-only action of the coupled system.
Section \ref{sec:results} is devoted to the analysis of this model
within a pertubative renormalization group and a self-consistent harmonic
approximation and discuss implications for the DC resistivity. Finally,
in Section \ref{sec:discussion} we discuss the main physical consequences
of our results and summarize them.

\section{\label{sec:model}Model}

We analyze the system depicted in Fig. \ref{fig:schema}, which represents
a clean superconducting wire of length $L$ and lateral dimensions
$d\ll\xi_{0}$, weakly coupled to a diffusive 2D metal. In the following
we use the convention $\hbar=k_{B}=1$. We begin our description with
the action of the microscopic BCS Hamiltonian for the isolated wire
\begin{align}
S_{\text{w}} & =\int_{0}^{\beta}d\tau\int d^{3}R\;\sum_{\sigma}\left\{ \bar{\psi}_{\sigma}\left(\partial_{\tau}-\mu\right)\psi_{\sigma}+\frac{\left[\nabla\bar{\psi}_{\sigma}\right]\left[\nabla\psi_{\sigma}\right]}{2m}\right\} \nonumber \\
 & +U\ \int_{0}^{\beta}d\tau\int d^{3}R\;\ \bar{\psi}_{\uparrow}\bar{\psi}_{\downarrow}\psi_{\downarrow}\psi_{\uparrow}.\label{eq:S_w}\end{align}
 Here the fermionic field $\psi_{\sigma}\equiv\psi_{\sigma}\left(\mathbf{R},\tau\right)$
describes an electron in the wire with spin projection $\sigma$ at
position $\mathbf{R}\equiv\left(x,y,z\right)$ and imaginary-time
$\tau$. The chemical potential $\mu=k_{F}^{2}/2m$ is the Fermi energy
in the normal state and the local attractive interaction $U<0$ is
responsible for pairing at $T<T_{c}$.

Assuming for simplicity that the coupling to the metallic film takes
place along the line $\left(x,0,0\right)$ in the wire, the coupling
term is described by

\begin{align}
S_{\perp} & =t_{\perp}\int_{0}^{\beta}d\tau\int dx\sum_{\sigma}\left[\bar{\psi}_{\sigma}\left(x,\tau\right)\eta_{\sigma}\left(x,\tau\right)+\text{h.c.}\right],\label{eq:S_perp}\end{align}
 where the fermionic field $\eta_{\sigma}\left(\mathbf{r},\tau\right)$
represents an electron at position $\mathbf{r}\equiv\left(x,y\right)$
in the film. Here the compact notation $\psi_{\sigma}\left(x,\tau\right)\equiv\left.\psi_{\sigma}\left(\mathbf{R},\tau\right)\right|_{y=z=0}$
, $\eta_{\sigma}\left(x,\tau\right)\equiv\left.\eta_{\sigma}\left(\mathbf{r},\tau\right)\right|_{y=0}$
has been used. While certainly more realistic models for the coupling,
which take into account geometrical details of the SN junction have
been studied \citep{hekking93, hekking94_andreev_conductance, feigelman00_keldysh_action_superconductor},
the main physics which is of interest to us is already captured by
Eq. (\ref{eq:S_perp}).

Electronic motion in the metallic film is described by the non-interacting
action

\begin{align}
S_{\text{2D}} & =\int_{0}^{\beta}d\tau\int d^{2}r\;\sum_{\sigma}\left\{ \bar{\eta}_{\sigma}\left(\partial_{\tau}-\mu_{\textrm{2D}}\right)\eta_{\sigma}+\right.\nonumber \\
 & \left.+\frac{\left[\nabla\bar{\eta}_{\sigma}\right]\left[\nabla\eta_{\sigma}\right]}{2m}+V\left(\mathbf{r}\right)\bar{\eta}_{\sigma}\left(\mathbf{r}\right)\eta_{\sigma}\left(\mathbf{r}\right)\right\} ,\label{eq:S_2D}\end{align}
 where $V\left(\mathbf{r}\right)$ is the (static) disorder potential.
We assume weak enough disorder, such that the localization length
in the film is $\xi_{\text{loc}}\gg L$, allowing us to neglect strong
localization effects.

For one given realization of the disorder potential $V\left(\mathbf{r}\right)$,
the effective action in the wire is obtained by integrating the electronic
degrees of freedom in the metallic film\begin{align*}
S_{\text{w}}^{\text{eff}} & =S_{\text{w}}+S_{\text{diss}}\\
S_{\text{diss}} & =-t_{\perp}^{2}\int_{0}^{\beta}d\tau d\tau^{\prime}\int dxdx^{\prime}\;\times\\
 & \times\sum_{\sigma}\bar{\psi}_{\sigma}\left(x,\tau\right)g_{\text{2D}}\left(x,x^{\prime};\tau-\tau^{\prime}\right)\psi_{\sigma}\left(x^{\prime},\tau^{\prime}\right),\end{align*}
 where $g_{\text{2D}}\left(\mathbf{r},\mathbf{r}^{\prime};\tau-\tau^{\prime}\right)$
is the Green's function in the film. Note that the spin index has
been dropped using the $SU\left(2\right)$ symmetry of the problem
and that we used the notation $g_{\text{2D}}\left(x,x^{\prime};\tau-\tau^{\prime}\right)\equiv\left.g_{\text{2D}}\left(\mathbf{r},\mathbf{r}^{\prime};\tau-\tau^{\prime}\right)\right|_{y=y^{\prime}=0}$.

Since the disorder potential breaks the original translation invariance
in the wire, an average over different configurations of the disorder
is needed to restore it. Let us define the partition function of the
systems for one disorder realization as

\begin{align*}
Z\left[V\right] & \equiv\int\mathcal{D}\left[\psi\right]\; e^{-S_{\text{w}}-S_{\text{diss}}},\end{align*}
 Assuming for convenience that $V\left(\mathbf{r}\right)$ is gaussian-distributed
\begin{align*}
S_{\textrm{d}} & =\frac{1}{2\mathcal{V}^{2}}\int d^{2}r\; V^{2}\left(\mathbf{r}\right),\end{align*}
 we can formally perform the average over different disorder configurations
as\begin{align*}
Z & =\frac{\int\mathcal{D}\left[V\right]\; e^{-S_{\text{d}}}Z\left[V\right]}{\int\mathcal{D}\left[V\right]\; e^{-S_{\text{d}}}}.\end{align*}
 Expansion of $Z\left[V\right]$ in powers of $t_{\perp}$ allows
us to obtain an explicit form of the partition function $Z$\begin{align}
Z & =\frac{\int\mathcal{D}\left[V\right]\; e^{-S_{\text{d}}}\int\mathcal{D}\left[\psi\right]\; e^{-S_{\text{w}}}\sum_{n=0}^{\infty}\frac{1}{n!}S_{\text{diss}}^{n}}{\int\mathcal{D}\left[V\right]\; e^{-S_{\text{d}}}},\nonumber \\
 & =\int\mathcal{D}\left[\psi\right]\; e^{-S_{\text{w}}}\sum_{n=0}^{\infty}\frac{1}{n!}\left\langle S_{\text{diss}}^{n}\right\rangle _{\text{d}}.\label{eq:Z_expansion}\end{align}

The low-energy effective action of this model is obtained by introducing
Hubbard-Stratonovich fields $\Delta\left(\mathbf{R},\tau\right),\Delta^{*}\left(\mathbf{R},\tau\right)$
in the particle-particle channel to decouple the quartic term in $S_{\text{w}}$
\citep{aitchison95_effective_bcs_lagragian, vanotterlo98, zaikin97,arutyunov08_superconductivity_1d_review}.
After integration of the fermionic degrees of freedom in the wire,
the action reads\begin{align}
S_{\text{w}}\left[\bar{\Delta},\Delta\right] & =-\text{Tr }\ln\mathbf{g}_{\text{w}}^{-1}-\frac{1}{U}\int_{0}^{\beta}d\tau\int d^{3}R\;\left|\Delta\left(\mathbf{R},\tau\right)\right|^{2},\label{eq:S_w_decoupled}\end{align}
 where \begin{align*}
\mathbf{g}_{\text{w}}^{-1} & \equiv\left[\begin{array}{cc}
\partial_{\tau}-\mu-\frac{\nabla^{2}}{2m} & \Delta\left(\mathbf{R},\tau\right)\\
\bar{\Delta}\left(\mathbf{R},\tau\right) & \partial_{\tau}+\mu+\frac{\nabla^{2}}{2m}\end{array}\right],\end{align*}
and where the Nambu notation \begin{align*}
\Psi\left(\mathbf{R},\tau\right) & =\left(\begin{array}{c}
\psi_{\uparrow}\left(\mathbf{R},\tau\right)\\
\bar{\psi}_{\downarrow}\left(\mathbf{R},\tau\right)\end{array}\right),\end{align*}
 is implicit. The Green's function in the wire formally reads \begin{align*}
\mathbf{g}_{\text{w}}\left(\mathbf{R},\tau\right) & \equiv\left[\begin{array}{cc}
g\left(\mathbf{R},\tau\right) & f\left(\mathbf{R},\tau\right)\\
\bar{f}\left(\mathbf{R},\tau\right) & \bar{g}\left(\mathbf{R},\tau\right)\end{array}\right],\end{align*}
where $g\left(\mathbf{R},\tau\right)\equiv\left\langle T_{\tau}\psi\left(\mathbf{R},\tau\right)\bar{\psi}\left(0\right)\right\rangle $
and $\bar{g}\left(\mathbf{R},\tau\right)\equiv\left\langle T_{\tau}\bar{\psi}\left(\mathbf{R},\tau\right)\psi\left(0\right)\right\rangle $
denote respectively the particle and hole propagators in the wire,
while $f\left(\mathbf{R},\tau\right)\equiv\left\langle T_{\tau}\psi\left(\mathbf{R},\tau\right)\psi\left(0\right)\right\rangle $,
$\bar{f}\left(\mathbf{R},\tau\right)\equiv\left\langle T_{\tau}\bar{\psi}\left(\mathbf{R},\tau\right)\bar{\psi}\left(0\right)\right\rangle $
are the anomalous ones\citep{fetter}.

For very narrow wires with diameter $d\ll\xi_{0}$ at low energies,
the dependence of the fields $\bar{\Delta}\left(\mathbf{R},\tau\right),\Delta\left(\mathbf{R},\tau\right)$
on transverse dimensions can be neglected, reducing to $\Delta\left(\mathbf{R},\tau\right)\rightarrow\Delta\left(\mathbf{x}\right)$
where the compact notation $\mathbf{x}=\left(x,\tau\right)$ has been
used. Moreover, at $T\ll T_{MF}$ (where $T_{MF}$ is the mean-field
critical temperature of Eq. (\ref{eq:S_w_decoupled})) and neglecting
amplitude fluctuations, the dynamical state of the wire is characterized
by $\Delta\left(\mathbf{x}\right)=\Delta_{0}e^{i\theta\left(\mathbf{x}\right)}$,
where the quantity $\Delta_{0}$ corresponds to the (temperature-dependent)
BCS energy-gap and $\theta\left(\mathbf{x}\right)$ is the space-
and time-dependent phase of the macroscopic BCS wavefunction.

The derivation of the phase-only action in the isolated wire (i.e.,
the first term in the expansion of Eq. (\ref{eq:Z_expansion})) is
obtained by the means of an expansion in Gaussian fluctuations in
the gradients of the field $\theta\left(\mathbf{x}\right)$ around
the BCS saddle-point, and takes the form of a Luttinger liquid action
\citep{giamarchi_book_1d, efetov74_fluctuations_1d_superconductor,aitchison95_effective_bcs_lagragian, zaikin97, arutyunov08_superconductivity_1d_review}
\begin{align}
S_{0} & =\int d\mathbf{x}\left[-i\Pi\partial_{\tau}\theta+\frac{uK}{2\pi}\left(\nabla\theta\right)^{2}+\frac{u}{2\pi K}\left(\pi\Pi\right)^{2}\right].\label{eq:S_ll}\end{align}
Here $\Pi\equiv\Pi\left(\mathbf{x}\right)$ is the momentum canonically
conjugate to $\theta\left(\mathbf{x}\right)$, formally defined through
the relation $\left[\theta\left(x\right),\Pi\left(x^{\prime}\right)\right]=i\delta\left(x-x^{\prime}\right)$
and representing fluctuations in the density of Cooper pairs at point
$\mathbf{x}$. The operator $\nabla$ denotes derivation with respect
to the spatial coordinate $x$. The Luttinger liquid parameters $u$
and $K$ are defined as \citep{zaikin97,giamarchi_book_1d}\begin{align*}
u & \equiv\sqrt{\frac{A_{\text{w}}n_{s}\left(T\right)}{4m\kappa\left(T\right)}},\\
K & \equiv2\pi\sqrt{\frac{A_{\text{w}}n_{s}\left(T\right)\kappa\left(T\right)}{4m}},\end{align*}
 where $\frac{n_{s}\left(T\right)}{4m}$ is the superconducting stiffness
of the wire (with $n_{s}\left(T\right)$ the 3D density of electrons
in the condensate and $m$ their mass), $A_{\text{w}}$ is the cross-sectional
area of the wire and $\kappa\left(T\right)$ is the compressibility
(cf. Ref. \onlinecite{zaikin97} for more details). $u$ corresponds
to the velocity of the plasma (Mooij-Sch{�}n\citep{mooij85_mooij_schon_mode})
mode. In the following, we assume the wire to be in the thermodynamic
limit $L\gg L_{T}=u/T$.

The next terms in the expansion of the partition function (Eq. (\ref{eq:Z_expansion}))
provide the effects of the coupling to the metallic film. At order
$\mathcal{O}\left(t_{\perp}^{2}\right)$ and at low temperatures $\left(T\ll T_{MF}\right)$,
the transfer of individual electrons is strongly forbidden by the
energy gap $\Delta_{0}$, giving a probability $\sim e^{-\Delta_{0}/T}$
for such a charge transfer channel.

The most relevant contribution of the coupling to the metallic film
appears at order $\mathcal{O}\left(t_{\perp}^{4}\right)$, and corresponds
to the Andreev reflection occurring at SN interfaces \citep{hekking93,hekking94_andreev_conductance,bruder94, feigelman98_smt_in_2d_proximity_array, feigelman00_keldysh_action_superconductor}.
This contribution physically represents processes in which \textit{paired}
electrons (for which there is no energy cost) are effectively transferred
from the wire to the film and viceversa \citep{bruder94}\begin{align}
S_{A} & =\frac{1}{2}\left\langle S_{\text{diss}}^{2}\right\rangle \nonumber \\
 & =2t_{\perp}^{4}\left[\int d\mathbf{x}^{\prime\prime}\; f\left(\mathbf{x}^{\prime\prime}\right)\right]^{2}\times\nonumber \\
 & \times\int d\mathbf{x}d\mathbf{x}^{\prime}P_{c}\left(\mathbf{x}-\mathbf{x}^{\prime}\right)\cos\left[\theta\left(\mathbf{x}\right)-\theta\left(\mathbf{x}^{\prime}\right)\right],\label{eq:S_diss_2}\end{align}
where $f\left(\mathbf{x}\right)=\left.f\left(\mathbf{R},\tau\right)\right|_{y=z=0}$.
The kernel $P_{c}\left(\mathbf{x}\right)=\left.P_{c}\left(\mathbf{r},\tau\right)\right|_{y=0}$
is the cooperon propagator in the diffusive film, defined as \citep{bruder94,akkermans}\begin{align}
P_{c}\left(\mathbf{r}-\mathbf{r}^{\prime},\tau-\tau^{\prime}\right) & \equiv\left\langle g_{\text{2D}}\left(\mathbf{r},\mathbf{r}^{\prime},\tau-\tau^{\prime}\right)g_{\text{2D}}\left(\mathbf{r},\mathbf{r}^{\prime},\tau-\tau^{\prime}\right)\right\rangle _{\text{d}},\label{eq:cooperon_def}\end{align}
representing the probability to find a coherent \textit{electron pair}
traveling a distance $\left|\mathbf{r}-\mathbf{r}^{\prime}\right|$
in the interval $\tau-\tau^{\prime}$ through the disordered film
\citep{akkermans} (see Fig. \ref{fig:schema}). The diffusive propagation
of this electron pair remains phase-coherent over a length $\xi_{N}$
(assumed $\gg\xi_{0}$) which depends on $T$, magnetic field and
the strength of Coulomb interactions \citep{akkermans}. In the absence
of the latter, $\xi_{N}\left(T\right)\simeq\sqrt{\frac{D}{T}}$, which
leads to important non-local coupling effects at low enough temperatures.

Explicit evaluation of Eq. (\ref{eq:cooperon_def}) for a diffusive
2D metal, assuming a Fermi-liquid description, yields (see Appendix
\ref{sec:cooperon}) \begin{align}
P_{c}\left(\mathbf{r},\tau\right) & \approx\frac{\rho_{\text{2D}}}{2\pi^{2}D}\tilde{P}_{c}\left(\mathbf{r},\tau\right),\label{eq:Pc_realspace}\end{align}
where we have defined \begin{align*}
\tilde{P}_{c}\left(\mathbf{r},\tau\right) & =\text{Re}\left\{ \frac{\exp\left(-\frac{r}{\xi_{N}}+\frac{ir^{2}}{4D\tilde{\tau}}\right)}{\tilde{\tau}^{2}}\Gamma\Bigl(0,\frac{ir^{2}}{4D\tilde{\tau}}\Bigr)\right\} .\end{align*}
Here $\Gamma\left(a,z\right)$ is the incomplete gamma function and
$\tilde{\tau}\equiv\tau+i\tau_{e}$, with $\tau_{e}$ the elastic
lifetime of electrons in the diffusive film \citep{akkermans}. Eq.
(\ref{eq:Pc_realspace}) is a valid expression for $\tau\gg\tau_{e}$
and $x\gg l_{e}$, where $l_{e}$ is the elastic mean-free path in
the film. In what follows, we set $y=0$ and consider the kernel as
depending only on the coordinate $\mathbf{x}$.

The coherence length $\xi_{N}\left(T\right)$ separates two distance
regimes of interest:

\textit{(a) the local regime} $x\gg\xi_{N}\left(T\right)$, where
the cooperon can be considered local in space, reducing to \begin{align*}
\tilde{P}_{c}\left(\mathbf{x}\right) & \approx\frac{\xi_{N}\delta\left(x\right)}{\tau^{2}}\ln\left(\frac{4D\tau}{\xi_{N}^{2}}\right),\end{align*}
consistent with the expression for the Andreev conductance in Refs.
\onlinecite{hekking93}. Introducing the notation $\mathbf{q}\equiv\left(k,\omega_{m}\right)$,
where $\omega_{m}$ is the bosonic Matsubara frequency $\omega_{m}=2\pi mT$,
the approximated Fourier transform (neglecting the logarithm) is independent
of $k$ for $k<\xi_{N}^{-1}$ and reads \begin{align*}
\tilde{P}_{c}\left(\mathbf{q}\right) & \approx2\xi_{N}\left[\frac{1}{\tau_{e}}-\pi\left|\omega_{m}\right|\right]\end{align*}
 in the limit $\mathbf{q}\rightarrow0$.

\textit{(b) The non-local regime} of distances $x<\xi_{N}\left(T\right)$,
where Eq. (\ref{eq:Pc_realspace}) can be approximated as\begin{align}
\tilde{P}_{c}\left(\mathbf{x}\right) & \approx\frac{\left(4D\right)^{2}}{x^{4}+\left(4D\tau\right)^{2}}\label{eq:Pc_approx}\end{align}
 with Fourier transform \begin{align*}
\tilde{P}_{c}\left(\mathbf{q}\right) & \approx2\pi^{2}\sqrt{D}\left[\sqrt{\frac{\pi}{\tau_{e}}}-2\sqrt{Dk^{2}+\left|\omega_{m}\right|}\right]\end{align*}
 for $\mathbf{q}\rightarrow0$.

It is convenient to introduce the normal state tunnel conductance
per unit of length in the SN junction \citep{tinkham}\begin{align*}
G_{t} & =\left(\frac{h}{2e^{2}}\right)\left(\frac{1}{2\pi}\right)^{2}t_{\perp}^{2}\frac{\rho_{\textrm{w}}\Omega_{\textrm{w}}\rho_{\text{2D}}A_{\textrm{2D}}}{L},\end{align*}
where $\rho_{\text{w}}\left(\rho_{\text{2D}}\right)$ is the normal-state
local density of states in the wire (film), and $\Omega_{\text{w}}$($A_{\text{2D}}$)
is the volume of the wire (area of the film). Replacing the expression
of the cooperon appearing in Eq. (\ref{eq:Pc_realspace}) and noting
that the resistivity in a 2D film is $\varrho_{2D}=\left[e^{2}\frac{n_{\textrm{2D}}}{m}\tau_{e}\right]^{-1}=\left[e^{2}\rho_{\textrm{2D}}D\right]^{-1}$,
we can express Eq. (\ref{eq:S_diss_2}) as\begin{align}
S_{A} & =\frac{G_{A}}{\xi_{0}^{2}}\int d\mathbf{x}d\mathbf{x}^{\prime}\tilde{P}_{c}\left(\mathbf{x}-\mathbf{x}^{\prime}\right)\cos\left[\theta\left(\mathbf{x}\right)-\theta\left(\mathbf{x}^{\prime}\right)\right],\label{eq:S_A}\end{align}
where $G_{A}$ is the dimensionless Andreev conductance in the SN
junction \citep{hekking93,hekking94_andreev_conductance} \begin{align*}
G_{A} & =\left(\frac{1}{2\pi}\right)^{4}4e^{2}G_{t}^{2}\varrho_{2D}.\end{align*}

In addition to the term $S_{A}$ of Eq. (\ref{eq:S_diss_2}), the
coupling $t_{\perp}$ generates contributions $\mathcal{O}\left(t_{\perp}^{2}\right)$
and $\mathcal{O}\left(t_{\perp}^{4}\right)$ at scales $x\lesssim\xi_{0}$
and $\tau\lesssim\xi_{0}u^{-1}$, which renormalize the bare Luttinger
parameters $K$ and $u$ of Eq. (\ref{eq:S_ll}) (e.g., diffuson propagator
\citep{akkermans}). Although these contributions do not change the
physics at a qualitative level, their effect is relevant for the comparison
with real systems. A microscopic study of the dependence of $K$ and
$u$ on the hopping $t_{\perp}$, as well as further renormalization
arising from Coulomb interactions between the wire and the film, is
beyond the scope of the present article and will be given elsewhere
\citep{lobos09}. In the following we assume that the Luttinger parameters
appearing in Eq. (\ref{eq:S_ll}) already include all these corrections.
Note also that the coupling to the metal modifies the bare value of
$\Delta_{0}$ through the well-known proximity-effect, in which the
diffusion of normal electrons in the superconductor produce a lowering
of $T_{c}$ \citep{DeGennes_supra}. However, since this is a small
effect of order $\mathcal{O}\left(t_{\perp}^{4}\right)$ and in addition
we assume $T\ll T_{c}$, this effect is irrelevant to our description
and can be effectively taken into account in renormalized values of
$\Delta_{0}$ and $T_{c}$.

So far we have not included the effects of topological defects (phase-slips)
in the wire. As discussed in Sec. \ref{sec:intro}, these topological
excitations produce finite resistivity at $T\ll T_{c}$ and are believed
to be the origin of destruction of superconductivity in narrow wires
\citep{zaikin97,arutyunov08_superconductivity_1d_review} and in dissipative
Josepshon junctions arrays \citep{goswami06_josephson_array,refael07_SN_transition_in_grains&nanowires}.
It can be shown \citep{giamarchi_book_1d} that defining a field $\phi\left(\mathbf{x}\right)$,
such that $\nabla\phi\left(\mathbf{x}\right)\equiv\pi\Pi\left(\mathbf{x}\right)$,
the generation of topological defects in the field $\theta\left(\mathbf{x}\right)$
can be described by a term \begin{align}
S_{\text{ps}} & =-\sum_{n=1}^{\infty}\frac{\lambda_{ps}^{n}u}{\xi_{0}^{2}}\int d\mathbf{x}\;\cos\left(2n\phi\left(\mathbf{x}\right)\right),\label{eq:S_ps}\end{align}
 where $\lambda_{ps}=\exp\left\{ -S_{\text{core}}\right\} $ is the
{}``fugacity'' of a phase-slip, and $S_{\text{core}}$ is the action
associated with the creation of a single phase-slip \citep{zaikin97,arutyunov08_superconductivity_1d_review}.
The term $\cos\left(2n\phi\left(\mathbf{x}\right)\right)$ represents
the creation of a kink of value $2\pi n$ in the $\theta$-field at
the space-time point $\mathbf{x}$. Assuming that $\lambda_{ps}\ll1$,
we can neglect in the following contributions with $n>1$ in $S_{\text{ps}}$.

Adding Eqs. (\ref{eq:S_ll}), (\ref{eq:S_A}) and (\ref{eq:S_ps})
we finally arrive at the expression of the effective phase-only action
at low temperatures \begin{alignat}{1}
S & =S_{0}+S_{A}+S_{\text{ps}},\label{eq:S_tot}\end{alignat}
describing on an equal footing the effects of fluctuation, dissipation
and topological excitations.

\section{\label{sec:results}Results}

\subsection{Renormalization-group analysis}

To study the properties of the model of Eq. (\ref{eq:S_tot}) at $T=0$,
we perform a renormalization-group analysis (RG) which is perturbative
in the couplings $G_{A}$ and $\lambda_{ps}$. At lowest possible
order, the RG equations are found by performing one- and two-loop
corrections in $S_{A}$ and $S_{ps}$, respectively. 

We adopt a renormalization procedure that rescales space and time
homogeneously, so as to preserve the Lorentz invariance of $S_{0}$. 

%
\begin{comment}
We define the ultraviolet cutoff of the theory to be $\tau_{c}=\max\left\{ \Delta_{0}^{-1},\tau_{e}\right\} $
and $x_{c}=\max\left\{ u,v_{F}\right\} \times\tau_{c}$.
\end{comment}
{}

The renormalization of $S_{A}$ involves a projection onto the most
relevant sector of the coupling kernel $\tilde{P}_{c}\left(\mathbf{x}\right)$,
which is very anisotropic in space and time, obeying a functional
RG-flow in the general case. We can simplify the analysis by studying
different scales of interest in the renormalization procedure. Depending
on the final scale $\Lambda\left(l\right)\sim L^{-1}$ (where $\Lambda\left(l\right)=\Lambda_{0}e^{l}$
is the renormalized momentum cut-off, and where $\Lambda_{0}^{-1}=\xi_{0}$),
we focus on the local part of the cooperon for $\Lambda\left(l\right)<\xi_{N}^{-1}$,
or on the non-local, diffusive properties for $\Lambda\left(l\right)>\xi_{N}^{-1}$.

We can motivate the RG analysis in the non-local regime by noting
that the kernel $\tilde{P}_{c}\left(\mathbf{x}\right)$ induce Josepshon
coupling of phases over spatial distances $W\left(\tau\right)\sim\sqrt{D\tau}$.
Indeed, an \textit{effective} purely local kernel $\tilde{P}_{c}^{\text{eff}}\left(\mathbf{x}\right)$
can be obtained by integrating the spatial coordinate in Eq. (\ref{eq:Pc_approx}),
yielding at long times \begin{align}
\tilde{P}_{c}^{\text{eff}}\left(\mathbf{x}\right) & \sim\tau^{-2}W\left(\tau\right)\delta\left(x\right)\quad\text{for }\tau\gg D/u^{2}.\label{eq:Pc_approx_eff}\end{align}
This approximate form is simpler to analyze, and yields a scaling
dimension $\frac{3}{2}$. Note that this long-range temporal $\tau^{-3/2}$
differs from the standard local ohmic coupling $\tau^{-2}$coupling
\citep{schoen90_review_ultrasmall_tunnel_junctions}, and further
quenches fluctuations of the phase. A more detailed functional RG
procedure involves an expansion of $\tilde{P}_{c}\left(\mathbf{x}\right)$
in terms of Legendre polynomials, and allows to extract the scaling
dimension of $\tilde{P}_{c}$ in the non-local limit in a systematic
way (cf. Appendix \ref{sec:cooperon}). Using that $\left\langle \cos\left(\theta\left(\mathbf{x}\right)-\theta\left(0\right)\right)\right\rangle \sim r^{-1/2K}$
for $r\rightarrow\infty$ (cf. Ref. \onlinecite{giamarchi_book_1d}),
we conclude that the scaling dimension of the perturbative term $S_{A}$
is $\frac{3}{2}-\frac{1}{2K}$.

In the local regime and for $\tilde{\lambda}_{ps}=0$, our the RG-analysis
reduces to that obtained in Ref. \onlinecite{cazalilla06_dissipative_transition},
where details of their derivation can be found. In this case, the
scaling analysis of $S_{A}$ is simpler to obtain, since in this limit
$\tilde{P}_{c}\left(\mathbf{x}\right)$ is RG-invariant, yielding
a scaling dimension of $1-\frac{1}{2K}$.

We obtain the flow equations

\begin{align}
\frac{dK\left(l\right)}{dl} & =\tilde{G}_{A}\left(l\right)-\tilde{\lambda}_{ps}^{2}\left(l\right)K^{3}\left(l\right)\label{eq:K_RG}\\
\frac{du\left(l\right)}{dl} & =\tilde{G}_{A}\left(l\right)\frac{u\left(l\right)}{K\left(l\right)}\frac{B^{\left(x\right)}-B^{\left(\tau\right)}}{B^{\left(x\right)}+B^{\left(\tau\right)}}.\label{eq:u_RG}\\
\frac{d\tilde{G}_{A}\left(l\right)}{dl} & =\left\{ \begin{array}{cc}
\left(1-\frac{1}{2K\left(l\right)}\right)\tilde{G}_{A}\left(l\right) & \text{(local)},\\
\left(\frac{3}{2}-\frac{1}{2K\left(l\right)}\right)\tilde{G}_{A}\left(l\right) & \text{(non-local)},\end{array}\right.\label{eq:lambdac_RG}\\
\frac{d\tilde{\lambda}_{ps}\left(l\right)}{dl} & =\left(2-K\left(l\right)\right)\tilde{\lambda}_{ps}\left(l\right),\label{eq:lambdaps_RG}\end{align}
where we have defined the dimensionless couplings $\tilde{G}_{A}\equiv G_{A}\pi\left(B^{\left(x\right)}+B^{\left(\tau\right)}\right)$
and $\tilde{\lambda}_{ps}\equiv\lambda_{ps}\sqrt{A}$ for convenience.
The dimensionless quantities $A$, $B^{\left(x,\tau\right)}$ are
non-universal and arise from the renormalization of $S_{ps}$ and
$S_{A}$, respectively, at scales $\left\{ x,u\tau\right\} <\Lambda^{-1}\left(l\right)$,
and are defined as 

\begin{align*}
A & \equiv\frac{1}{4\pi}\int_{\Lambda^{-1}\left(l\right)}^{\infty}d\tilde{r}\;\tilde{r}^{3}e^{-2KF_{\Lambda}\left(\tilde{r}\right)}F_{\Lambda}\left(\tilde{r}\right),\\
B^{\left(x\right)} & \equiv\frac{1}{2}\frac{1}{\Lambda^{2}\left(l\right)u^{2}}\int_{0}^{2\pi}d\phi\;\tilde{P}_{c}\left(\Lambda^{-1}\left(l\right),\phi\right)\cos^{2}\phi,\\
B^{\left(\tau\right)} & \equiv\frac{1}{2}\frac{1}{\Lambda^{2}\left(l\right)u^{2}}\int_{0}^{2\pi}d\phi\;\tilde{P}_{c}\left(\Lambda^{-1}\left(l\right),\phi\right)\sin^{2}\phi,\end{align*}
 where $\tilde{P}_{c}$ has been expressed in cylindrical coordinates
(cf. Eq. \ref{eq:Pc_polar}), and where $\tilde{r}\equiv\Lambda\left(l\right)\sqrt{x^{2}+\left(u\tau\right)^{2}}$
and $F_{\Lambda}\left(\mathbf{x}\right)=\frac{1}{2}\ln\tilde{r}$.
It is interesting to point out that while only one parameter $A$
arises in the rescaling of $S_{ps}$ due to space-time isotropy, the
anisotropy of $S_{A}$ generates different parameters $B^{\left(x\right)}$and
$B^{\left(\tau\right)}$.

Note that in the local regime ($\tilde{P}_{c}\sim\tilde{P}_{c}\left(\tau\right)$),
the product $uK$ does not renormalize for $\lambda_{ps}=0$, and
thus $B^{\left(x\right)}=0$. Non-locality is thus captured by a $B^{\left(x\right)}>0$.
Further, since the term $S_{A}$ breaks the space-time isotropy within
our Lorentz-invariant RG analysis (i.e., momentum shell integration
homogeneous in space-time), we expect a renormalization of the velocity
$u\left(l\right)$, cf. Eq. (\ref{eq:u_RG}). A numerical evaluation
gives $B^{\left(x\right)}/B^{\left(\tau\right)}<1$, meaning that
$u\left(l\right)$ flows towards smaller values.

As for Eq. (\ref{eq:lambdaps_RG}), we note that it corresponds to
the usual Berezinskii-Kosterlitz-Thouless (BKT) flow equation (cf.
Ref. \onlinecite{zaikin97} for a derivation in the context of superconducting
wires).

In the limit $\left\{ \tilde{G}_{A}\left(l\right),\tilde{\lambda}_{ps}\left(l\right)\right\} \rightarrow0$,
the properties of the system are dominated by the value of $K\left(l\right)$.
From Eqs. (\ref{eq:lambdac_RG}) and (\ref{eq:lambdaps_RG}), we can
define the critical values $K_{A}^{*}\equiv\frac{1}{2}\left(\equiv\frac{1}{3}\right)$
for the local (non-local) regime, and $K_{ps}^{*}\equiv2$. For $\tilde{\lambda}_{ps}=0$
and $K>K_{A}^{*}$, the coupling $\tilde{G}_{A}\left(l\right)$ flows
towards strong coupling and eventually the perturbative RG analysis
is no longer valid. On the other hand, for $\tilde{G}_{A}=0$ and
$K<K_{ps}^{*}$ the coupling $\tilde{\lambda}_{ps}\left(l\right)$,
becomes relevant and eventually superconductivity is destroyed in
the wire, due to the unbinding of pairs of topological excitations
\citep{giamarchi_book_1d,zaikin97}. Note that it is not possible
to determine the nature of the $T=0$ fixed point in this regime within
our formalism. This issue is currently under intensive research \citep{arutyunov08_superconductivity_1d_review}.

Therefore, at $T=0$ and when neither $\tilde{G}_{A}$ nor $\tilde{\lambda}_{ps}$
vanish, the Luttinger liquid phase is never stable, and the ground
state is determined by a competition between $S_{A}$ and $S_{ps}$.

\subsection{Self-consistent Harmonic Approximation }

To further investigate the properties in the regime where $\tilde{G}_{A}$
is the dominant parameter that flows to strong coupling, we used a
self-consistent variational approach, the so-called self-consistent
harmonic approximation \citep{feynman_statmech,giamarchi_book_1d}.
This method consists in finding the optimal propagator $g_{\text{tr}}\left(\mathbf{q}\right)$
of a harmonic (Gaussian) trial action \begin{align*}
S_{\text{tr}}\left[\theta\right] & =\frac{1}{2\beta L}\sum_{\mathbf{q}}\frac{1}{g_{\text{tr}}\left(\mathbf{q}\right)}\left|\theta\left(\mathbf{q}\right)\right|^{2},\end{align*}
 that minimizes the variational free-energy \begin{align*}
F_{\text{var}} & =F_{\text{tr}}+T\left\langle S-S_{\text{tr}}\right\rangle _{\text{tr}},\end{align*}
 where \begin{align*}
F_{\text{tr}} & =-T\:\ln\int\mathcal{D}\theta\; e^{-S_{\text{tr}}\left[\theta\right]}.\end{align*}
 The minimization of the free-energy $F_{\text{var}}$ with respect
to $g_{\text{tr}}\left(\mathbf{q}\right)$ yields a self-consistent
equation for $g_{\text{tr}}\left(\mathbf{q}\right)$

\begin{align}
g_{\text{tr}}^{-1}\left(\mathbf{q}\right) & =g_{0}^{-1}\left(\mathbf{q}\right)-\frac{2G_{A}}{\xi_{0}^{2}}\int d\mathbf{x}\;\left[\cos\left(\mathbf{q}\mathbf{x}\right)-1\right]\times\nonumber \\
 & \times\tilde{P}_{c}\left(\mathbf{x}\right)\exp\left\{ -\frac{1}{\beta L}\sum_{\mathbf{q}^{\prime}}\left[1-\cos\left(\mathbf{q}^{\prime}\mathbf{x}\right)\right]g_{\text{tr}}\left(\mathbf{q}^{\prime}\right)\right\} ,\label{eq:g_tr_equation}\end{align}
where $g_{0}^{-1}\left(\mathbf{q}\right)\equiv\frac{K}{\pi u}\omega_{m}^{2}+\frac{uK}{\pi}k^{2}$
is the propagator in the Luttinger liquid phase. The solutions of
Eq. (\ref{eq:g_tr_equation}) read\begin{align}
g_{\text{tr}}^{-1}\left(\mathbf{q}\right) & =\left\{ \begin{array}{cc}
g_{0}^{-1}\left(\mathbf{q}\right)+\eta\left|\omega_{m}\right| & \text{(local)},\\
g_{0}^{-1}\left(\mathbf{q}\right)+\eta\sqrt{Dk^{2}+\left|\omega_{m}\right|} & \text{(non-local)}.\end{array}\right.\label{eq:g0_diffusive}\end{align}
 The parameter $\eta$ is found self-consistently for the general
case, but in the limit $G_{A}\rightarrow0$ it reduces to

\begin{align*}
\eta & =\begin{cases}
{\displaystyle \left[\frac{2\pi G_{A}\xi_{N}\exp\left(\frac{\gamma}{2K}\right)}{\xi_{0}^{2}}\right]^{\frac{2K}{2K-1}}\left[\frac{\pi\xi_{0}}{4K}\right]^{\frac{1}{2K-1}}} & \text{(local)},\\
{\displaystyle \left[\frac{8\pi^{2}G_{A}\sqrt{D}}{\xi_{0}^{2}}\right]^{\frac{3K}{3K-1}}\left[\frac{\pi\xi_{0}^{3}u}{4K\sqrt{D^{3}}}\right]^{\frac{1}{3K-1}}} & \text{(non-local)}.\end{cases}\end{align*}
Note that physical solutions of the Eq. (\ref{eq:g_tr_equation})
with $\eta\neq0$ are found only for $K>K_{A}^{*}$, confirming the
results of the RG analysis. In the context of the variational approach,
it becomes clear (cf. Eq. (\ref{eq:g0_diffusive})) that the contribution
of the cooperon propagator of Eq. (\ref{eq:S_A}) induces ohmic (non-ohmic)
dissipation in the local (non-local) regime. Evaluation of the phase-correlation
function at $T=0$ with the optimal $g_{\text{tr}}\left(\mathbf{q}\right)$
of Eq. (\ref{eq:g0_diffusive}), yields in the long wavelength limit\begin{align}
\frac{\left\langle e^{i\theta\left(\mathbf{x}\right)-i\theta\left(0\right)}\right\rangle }{\left\langle e^{i\theta}\right\rangle ^{2}} & \simeq\begin{cases}
{\displaystyle 1+\frac{1}{\sqrt{\pi}\eta}\frac{1}{x+\sqrt{\frac{8uK}{\pi\eta}\tau}}} & \text{(local)},\\
{\displaystyle 1+\frac{2\sqrt{D}}{\eta\pi^{2}}\frac{1}{x^{2}+4D\tau}} & \text{(non-local)},\end{cases}\label{eq:corr_scha}\end{align}
 where $\left\langle e^{i\theta}\right\rangle $ is the value of the
superconducting order parameter\begin{align}
\left\langle e^{i\theta}\right\rangle  & =\begin{cases}
{\displaystyle \left[\frac{\pi\xi_{0}\eta}{4K}\right]^{1/4K}} & \text{(local)},\\
{\displaystyle \left[\frac{\pi\xi_{0}^{3}u\eta}{4K\sqrt{D^{3}}}\right]^{1/6K}} & \text{(non-local)}.\end{cases}\label{eq:order_parameter_scha}\end{align}
 This result indicates that the order parameter develops long-range
order, and should be compared with the case of isolated wires, where
superconducting correlation functions follow a power-law behavior
and $\left\langle e^{i\theta}\right\rangle =0$ as a consequence of
the strong quantum fluctuations \citep{mermin_wagner_theorem}.

\subsection{DC transport properties }

Now we address the experimentally relevant question of the possibility
to observe some signatures of our predictions at $T=0$. To that end,
we turn our attention to transport properties, and calculate the DC
resistivity. We use the theoretical framework of the memory matrix,
which is perturbative in the processes that degrade the current-density
operator \citep{gotze_fonction_memoire,giamarchi_attract_1d,giamarchi_book_1d}
\begin{align*}
J\left(\mathbf{x}\right) & =\frac{uK}{\pi}\frac{2e}{c}\nabla\theta\left(\mathbf{x}\right).\end{align*}
Current-decay originated by phase-slips induce finite resisitivity
at $T<T_{c}$. At very low temperatures $T\ll T_{c}$ thermally activated
phase slips are suppressed and resistivity is dominated by quantum
phase-slips processes. In the absence of dissipation ($G_{A}=0$),
its expression is well-known and reads \citep{giamarchi_attract_1d, zaikin97}\begin{align}
\varrho\left(T\right) & \approx\frac{4\pi^{3}\tilde{\lambda}_{ps}^{2}\Lambda_{0}}{\left(\frac{2e}{c}\right)^{2}}B^{2}\left(\frac{K}{2},1-K\right)\cos^{2}\bigl(\frac{\pi K}{2}\bigr)\bigl(\frac{2\pi T}{u\Lambda_{0}}\bigr)^{2K-3},\label{eq:resistivity}\end{align}
where $B\left(x,y\right)$ is the beta function. This is a valid expression
provided that a perturbation expansion in $\lambda_{ps}$ and $G_{A}$
is possible. At finite temperatures, the effect of these couplings
can be incorporated by replacing the bare parameters by the renormalized
ones obtained from the integration of the RG-flow equations (Eqs.
(\ref{eq:K_RG}-\ref{eq:lambdaps_RG})) up to a scale\citep{giamarchi_attract_1d}
$\Lambda^{-1}\left(l\right)=u\left(l\right)/2\pi T$ . 

\begin{figure}
\includegraphics[bb=0bp 0bp 200bp 130bp,clip,scale=1.35]{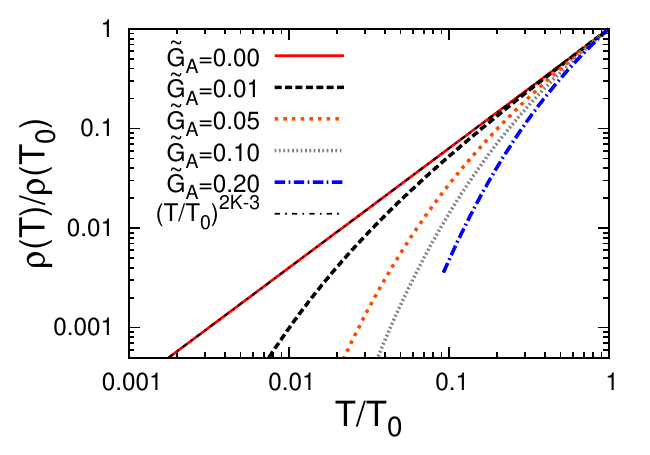}

\caption{(color online) Normalized resistivity vs $T/T_{0}$. As the (dimensionless)
Andreev conductance $\tilde{G}_{A}$ is increased, the wire resistivity
$\varrho$$\left(T\right)$ deviates from the law $\sim T^{2K-3}$
predicted for an isolated wire \citep{giamarchi_attract_1d, zaikin97, arutyunov08_superconductivity_1d_review}
as a consequence of the dissipation-induced increase of the stiffness
$K$ (cf. Eq. (\ref{eq:K_RG})). \label{fig:resistivity}}

\end{figure}

Our results are shown in Fig. \ref{fig:resistivity}, where we calculate
$\varrho\left(T\right)$ normalized to the {}``high-temperature''
value $T_{0}=\xi_{0}/u$, fixed by the short time cut-off of the theory.
According with our estimations (see Sec. \ref{sec:discussion}), we
analyze only the local regime $\Lambda\left(l\right)\ll\xi_{N}^{-1}$.
We start with the initial conditions $K\left(0\right)=2.1,\tilde{\lambda}_{ps}\left(0\right)=10^{-3}$,
for $\tilde{G}_{A}\left(0\right)=0$ (solid line in Fig. \ref{fig:resistivity}).
For comparison, we show the $\left(T/T_{0}\right)^{2K\left(0\right)-3}$
behavior predicted for the resistivity due to phase-slips in the absence
of dissipation effects (dot-dashed line)\citep{giamarchi_attract_1d, zaikin97, arutyunov08_superconductivity_1d_review}. 

Interestingly, starting the RG flow with the initial values $\tilde{G}_{A}\left(0\right)=0.01,\;0.05,\;0.1\text{ and }0.2$,
the resistivity decreases \textit{faster} than the $\left(T/T_{0}\right)^{2K\left(0\right)-3}$
law corresponding to the isolated wire. This illustrates the stabilizing
effect of dissipation on superconductivity, which manifests itself
through an increase of the stiffness $K$, as can be seen from Eq.
(\ref{eq:K_RG}) when parameter $\tilde{G}_{A}$ dominates over $\tilde{\lambda}_{ps}$. 

Note that since the integration of the renormalization group flow
(and consequently, the calculation of the resistivity) is perturbative,
it cannot be carried beyond a point where either $\tilde{G}_{A}\left(l\right)$
or $\tilde{\lambda}_{ps}\left(l\right)$ become of order unity.

\section{\label{sec:discussion}Discussion and summary}

The result of the RG flow Eqs. (\ref{eq:K_RG})-(\ref{eq:lambdaps_RG})
together with the analysis in the strong coupling regime, summarized
in Eqs. (\ref{eq:corr_scha}) and (\ref{eq:order_parameter_scha}),
suggest that a weak coupling to the metallic film favors a superconducting
ground-state with long-range order of the order parameter at $T=0$,
through a dissipation-induced quench of phase fluctuations. 

Note that this is not trivial, since a \textit{strong} coupling to
a disordered metallic film is detrimental to superconductivity and
lowers $T_{c}$ through the well-known proximity effect \citep{DeGennes_supra}.
But in a low-dimensional situation at $T\ll T_{c}$, where phase fluctuations
are the dominant mechanism of destruction of global phase-coherence,
the environment actually favors long-range order. 

This picture is supported by experiments on disordered granular Pb
films coated with a thin Ag metallic film \citep{merchant01_crossover_phase_amplitude_fluctuations},
where it was shown that while $T_{c}$ \textit{decreases} due to the
proximity effect, phase stiffness actually \textit{increases} at low
enough temperatures. Also in the context of dissipative Josephson
junctions arrays, it is well-known that the existence of coupling
to a normal metal stabilizes the superconducting phases \citep{schoen90_review_ultrasmall_tunnel_junctions,fazio01_review_superconducting_networks,  bobbert92_transitions_dissipative_josephson_chains,goswami06_josephson_array,refael07_SN_transition_in_grains&nanowires}. 

A similar idea has been recently suggested to produce an enhancement
of $T_{c}$ in bilayered materials \citep{berg08_route_to_htcs_composite_systems},
in which one layer has a high pairing scale but low superfluid density
while in the other layer the situation is the inverse. When both materials
are put into contact, the $T_{c}$ of the coupled system is higher
than those of the isolated layers. In the specific case of Luttinger
liquids coupled to dissipative baths, our results are in agreement
with recent theoretical works where the existence of superconductive
long-range order at $T=0$ has been suggested \citep{cazalilla06_dissipative_transition,artemenko07_longrange_order_in_1d}. 

In this paper, we have presented a rigorous study of a realistic dissipative
mechanism, provided by a coupling to a diffusive metal, in the context
of superconducting wires. 

Of central importance in our analysis is the cooperon propagator kernel
$\tilde{P}_{c}\left(\mathbf{x}-\mathbf{x}^{\prime}\right)$, which
couples the field $\theta$ at the space-time coordinates $\mathbf{x}$
and $\mathbf{x}^{\prime}$ in the dissipative term $S_{A}$ (cf. Eq.
(\ref{eq:S_A})). The physics of the kernel $\tilde{P}_{c}\left(\mathbf{x}-\mathbf{x}^{\prime}\right)$
strongly depends on the relation between $\xi_{N}\left(T\right)$,
the coherence length in the diffusive film, and the length of the
wire $L$. Consequently, two regimes of interest appear: the local
regime $\xi_{N}\left(T\right)\ll L$, where the coupling of phases
is purely local in space and non-local in time; and the non-local
regime $\xi_{N}\left(T\right)>L$, where the phase-coupling is non-local
both in space and time. At this point, it is interesting to determine
to which regime actual superconducting wires would correspond. At
the experimentally relevant temperature $T\simeq1K$, and using typical
values of $D$ in clean metallic films\citep{morpurgo09_private_communication}
$D\sim10^{2}\text{cm}.\text{s}^{-1}$, we obtain the estimate $\xi_{N}\left(1K\right)\sim0.1\;\mu m$.
%
\begin{comment}
\begin{align*}
\xi_{N} & =\sqrt{\frac{\hbar D}{2\pi k_{B}T_{c}}}\\
 & =\sqrt{\frac{1.0\times10^{-34}J.s\;10^{-2}m^{2}s^{-1}}{1.0\times10^{-23}J.K^{-1}\times1K}}\\
 & =\sqrt{\frac{1.0\times10^{-13}m^{2}}{1.0}}\\
 & \approx0.1\;\mu m\end{align*}

\end{comment}
{}On the other hand, the temperature constraints to observe non-local
effects can be compactly written as $L_{T}\left(T\right)\ll L\ll\xi_{N}\left(T\right)$.
These conditions require that $T\gg T_{NL}\equiv\hbar u^{2}/k_{B}D$
(where units have been restored) while the length has to be kept smaller
than $\xi_{N}\left(T\right)$. Estimating the velocity of the Mooij-Sch{\"o}n
as $u\sim10^{5}\text{m.s}^{-1}$ for the wires in Ref. \onlinecite{giordano88_evidence_of_qps},
we obtain $T_{NL}\sim10-100\; K$, which exceeds the bulk-$T_{c}$
values estimated in the range $6.9-7.1\; K$ in In-Pb films\citep{giordano94}.
The above estimates show that spatial non-local effects are elusive
in actual wires (e.g., such as those studied in Ref. \onlinecite{altomare06_experimental_evidence_of_qps},
where $L\sim10-100\;\mu m$), but may eventually be observed in superconducting
wires with higher $T_{c}$, coupled to very clean substrates. 

In order to make contact with recent transport experiments\citep{giordano88_evidence_of_qps,giordano94,bezryadin00,lau01,altomare06_experimental_evidence_of_qps},
we have calculated the linear DC resistivity of a wire weakly coupled
to a diffusive film, for different values of the Andreev conductance
$G_{A}$. The results of Fig. \ref{fig:resistivity}, calculated for
the local regime $\xi_{N}\left(T\right)\ll L$, suggest that signatures
of the predicted long-range order phase at $T=0$ could be observed
experimentally. Indeed, since the dissipative term $S_{A}$ renormalizes
the superconducting stiffness $K$ to higher values as the temperature
decreases (cf. Eq. (\ref{eq:K_RG})), sizable deviations from the
predictions for an isolated wire\citep{giamarchi_attract_1d, zaikin97}
$\varrho\left(T\right)\sim T^{2K\left(0\right)-3}$, where $K\left(0\right)$
is the bare stiffness, could be achieved at low enough temperatures.

%
\begin{comment}
In this section we analyze the relation of our theoretical description
with an experimental situation.

First we translate our parameters to experimental concepts.

Our predictions apply to wires which are in the thermodynamical regime
$L>u\beta$. This fact restricts the range of temperatures to \begin{align*}
T_{L}< & T & \ll T_{c}\end{align*}
 where $T_{L}\equiv u/L$. Taking as a concrete example the wires
in Ref. \citet{giordano94} with typical lengths $\sim100\mu m$ and
$d\sim10nm$, and estimating $u$ from \begin{align*}
u & \sim c\frac{d}{\lambda_{L}}\sim3\times10^{8}\frac{m}{s}\times\frac{10\times10^{-9}m}{130\times10^{-9}m}\\
 & \sim3.10^{7}\frac{m}{s}.\end{align*}
 Then, for $L\sim100\mu m$, \begin{eqnarray*}
T_{L} & = & \frac{u\hbar}{k_{B}L}\\
 & \sim & \frac{10^{7}m/s}{10^{-4}m}.\frac{10^{-34}J.s}{10^{-23}J/K}\\
 & \sim & 1K.\end{eqnarray*}

\end{comment}
{}{}%
\begin{comment}
\begin{eqnarray*}
L_{T} & = & \frac{u\hbar}{k_{B}T}\\
 & = & \frac{3.10^{7}m/s}{3K}.\frac{10^{-34}J.s}{10^{-23}J/K}\\
 & = & 100\mu m\end{eqnarray*}

\end{comment}
{}{}

In summary, we have studied a thin superconducting wire weakly coupled
to a metallic film, focusing on the details of dissipation provided
by the metallic cooperons at low temperatures. We have studied the
phase diagram at $T=0$ within the framework of renormalization group
and a variational harmonic approximation. We predict a quantum phase
transition towards a superconductor with long-range order at $T=0$
as a function of the Andreev conductance $G_{A}$ and the superconducting
stiffness $K$ of the wire. Finally, we show that some signatures
of this ordered phase could be observed in experiments of transport,
manifested as an increase of the superconducting stiffness and consequently
the exponent of $\varrho\left(T\right)\sim T^{\nu}$ at low temperatures.

\begin{acknowledgments}
The authors are grateful to L. Benfatto and M. A. Cazalilla for useful
discussions. This work was partially supported by the Swiss National
Foundation under MaNEP and division II and by CONICET and UNLP. 
\end{acknowledgments}
\appendix

\section{\label{sec:cooperon}The cooperon propagator}

In this section we derive the expression for the cooperon propagator
assuming weak disorder in the metallic film. We refer the reader to
Ref. \onlinecite{akkermans} for further details. 

When evaluating averages over the disorder potential in Eq. (\ref{eq:cooperon_def})
\begin{align*}
P_{c}\left(\mathbf{r}-\mathbf{r}^{\prime},\tau-\tau^{\prime}\right) & \equiv\left\langle g_{\text{2D}}\left(\mathbf{r},\mathbf{r}^{\prime};\tau-\tau^{\prime}\right)g_{\text{2D}}\left(\mathbf{r},\mathbf{r}^{\prime};\tau-\tau^{\prime}\right)\right\rangle _{\text{d}}\end{align*}
 a diagrammatic series (ladder diagrams) is constructed upon the repeated
action of the Dyson's equation (in operator notation)\begin{align*}
\hat{\mathbf{g}}_{\text{2D}}= & \hat{\mathbf{g}}_{\text{2D}}^{0}+\hat{\mathbf{g}}_{\text{2D}}^{0}\hat{\mathbf{V}}\hat{\mathbf{g}}_{\text{2D}}\end{align*}
where $\hat{\mathbf{g}}_{\text{2D}}^{0}$ is the unperturbed electron
Green's function in the otherwise perfect metal, and $\delta\left(\mathbf{r}-\mathbf{r}^{\prime}\right)V\left(\mathbf{r}\right)=\left\langle \mathbf{r}\left|\hat{\mathbf{V}}\right|\mathbf{r}^{\prime}\right\rangle $
is the (static) disorder potential which verifies

\begin{align}
\left\langle V\left(\mathbf{r}\right)\right\rangle _{\text{d}} & =0\label{eq:U_average}\\
\left\langle V\left(\mathbf{r}\right)V\left(\mathbf{r}^{\prime}\right)\right\rangle _{\text{d}} & =n_{i}\mathcal{V}^{2}\delta\left(\mathbf{r}-\mathbf{r}^{\prime}\right)\label{eq:U_correlation}\end{align}
where $n_{i}$ is the concentration of impurities and $\mathcal{V}$
is the uniform component of $V\left(\mathbf{r}\right)$ in Fourier
space. The diagrammatic series in Fourier space representation is
given by (cf. Ref. \onlinecite{akkermans}) \begin{align*}
P_{c}\left(\mathbf{Q};n,m\right) & =\frac{P_{c}^{0}\left(\mathbf{Q};n,m\right)}{1-\frac{n_{i}\mathcal{V}^{2}}{\Omega}P_{c}^{0}\left(\mathbf{Q};n,m\right)},\end{align*}
where \begin{align*}
P_{c}^{0}\left(\mathbf{Q};n,m\right) & \equiv\sum_{\mathbf{p}}g_{\text{2D}}\left(\mathbf{p},i\nu_{n}\right)g_{\text{2D}}\left(\mathbf{Q}-\mathbf{p},i\nu_{m}\right),\end{align*}
and where $\Omega$ is the volume of the sample. $\mathbf{Q}=\mathbf{k}+\mathbf{k}^{\prime}$
represents the center-of-mass momentum of the two electron system,
and $\mathbf{k}$ and $\mathbf{k}^{\prime}$ are the initial (i.e.,
before colliding with impurities) momenta of the individual electrons.
Defining\begin{align*}
i\omega_{l} & \equiv i\nu_{m}-i\nu_{n},\\
\zeta\left(\mathbf{Q};i\nu_{n},i\omega_{l}\right) & \equiv\frac{n_{i}\mathcal{V}^{2}}{\Omega}P_{c}^{0}\left(\mathbf{Q};n,l+n\right),\end{align*}
the real-space representation of $P_{c}$ reads

\begin{align}
P_{c}\left(\mathbf{r},\tau\right) & =\left(\frac{1}{\beta\Omega}\right)^{2}\sum_{\mathbf{Q=k+k^{\prime}}}\sum_{n,l}e^{-i\left(2\nu_{n}+\omega_{l}\right)\tau}e^{i\mathbf{Q}.\mathbf{r}}\times\nonumber \\
 & \times\frac{\frac{\Omega}{n_{i}\mathcal{V}^{2}}\zeta\left(\mathbf{Q};i\nu_{n},i\omega_{l}\right)}{1-\zeta\left(\mathbf{Q};i\nu_{n},i\omega_{l}\right)}.\label{eq:ladder}\end{align}
Using the definition of the mean-free path $l_{e}\equiv v_{F}\tau_{e}$,
and the definition of the elastic lifetime (in Born's approximation)
\citep{akkermans}

\begin{align}
\frac{1}{\tau_{e}} & \equiv2\pi\rho_{\text{2D}}n_{i}\mathcal{V}^{2}\label{eq:tau_e}\end{align}
 we obtain in the diffusive limit $\left(Ql_{e},\omega_{l}\tau_{e}\ll1\right)$
\begin{align*}
\zeta\left(Q;i\nu_{n},i\omega_{l}\right) & \approx1-\tau_{e}\left(\left|\omega_{l}\right|+DQ^{2}\right),\end{align*}
 where we have defined the diffussion constant in $d$ spatial dimensions
$D\equiv\frac{v_{F}^{2}\tau_{e}}{d}=\frac{l_{e}^{2}}{\tau_{e}d}$.
Replacing these results into Eq. (\ref{eq:ladder}), and noting that
the contribution to $\zeta\left(\mathbf{Q};i\nu_{n},i\omega_{l}\right)$
is vanishingly small for $\nu_{n}\left(\nu_{n}+\omega_{l}\right)>0$,
we obtain\begin{align}
P_{c}\left(\mathbf{r},\tau\right) & \approx2\pi\rho_{\text{2D}}\frac{1}{\beta}\sum_{\nu_{n}}e^{-i2\nu_{n}\tau}\times.\nonumber \\
 & \times\left.\frac{1}{\beta\Omega}\sum_{\omega_{l},\mathbf{Q}}\frac{e^{i\mathbf{Q}.\mathbf{r}}e^{-i\omega_{l}\tau}}{\left|\omega_{l}\right|+DQ^{2}}\right|_{\nu_{n}\left(\nu_{n}+\omega_{l}\right)<0}\label{eq:true_coop}\end{align}
In addition, and since the condition $\nu_{n}\left(\nu_{n}+\omega_{l}\right)<0$
must be fulfilled, we note that the limit $\left|\omega_{l}\right|\tau_{e}\rightarrow0$
constrains the summation over Matsubara frequencies $\nu_{n}$ to
values near $\nu_{n}\approx0$. Therefore, at low temperatures we
find\begin{align*}
P_{c}\left(\mathbf{r},\tau\right) & =2\rho_{\text{2D}}\frac{1}{\beta\Omega}\sum_{\omega_{l},\mathbf{Q}}\frac{\sin\left(\left|\omega_{l}\right|\tau\right)}{\tau}\frac{e^{i\mathbf{Q}.\mathbf{r}}}{\left|\omega_{l}\right|+DQ^{2}}.\end{align*}
We can generalize this expression to take into account processes that
break phase coherence (magnetic fields, magnetic impurities, etc)
\begin{align}
P_{c}\left(\mathbf{r},\tau\right) & =2\rho_{\text{2D}}\frac{1}{\beta\Omega}\sum_{\omega_{l},\mathbf{Q}}\frac{\sin\left(\left|\omega_{l}\right|\tau\right)}{\tau}\frac{e^{i\mathbf{Q}.\mathbf{r}}}{\left|\omega_{l}\right|+DQ^{2}+\tau_{\varphi}^{-1}},\label{eq:cooperon}\end{align}
 where $\tau_{\varphi}$ is a phenomenological phase-breaking time.
At $T=0$ we can replace $\frac{1}{\beta}\sum_{\omega_{l}}\rightarrow\frac{1}{2\pi}\int_{0}^{\infty}d\omega.$
Then, the sum over $\mathbf{Q}$ in Eq. (\ref{eq:cooperon}) yields
for a 2D system\begin{align*}
P_{c}\left(\mathbf{r},\tau\right) & =\text{Re}\left\{ \frac{\rho_{\text{2D}}}{i\tau\pi^{2}D}\int_{0}^{\infty}d\omega\; e^{\omega\left(i\tau-\tau_{c}\right)}\times\right.\\
 & \left.\times K_{0}\left(\sqrt{\frac{\omega+\tau_{\varphi}^{-1}}{D}}r\right)\right\} ,\end{align*}
where $K_{0}\left(x\right)$ is the zeroth-order modified Bessel function.
$r=\sqrt{x^{2}+y^{2}}$ is the distance in the $x-y$ plane and $\tau_{c}$
is an ultraviolet cutoff in time to make the integral in $\omega$
convergent, and which we set $\tau_{c}=\tau_{e}$. Finally we arrive
at the expression of the cooperon in Eq. (\ref{eq:Pc_realspace})
\begin{align}
P_{c}\left(\mathbf{r},\tau\right) & =\frac{\rho_{\text{2D}}}{2\pi^{2}D}\tilde{P}_{c}\left(\mathbf{r},\tau\right),\end{align}
where we defined \begin{align*}
\tilde{P}_{c}\left(\mathbf{r},\tau\right) & =\text{Re}\left\{ \frac{ie^{-\frac{r}{\xi_{N}}}e^{\frac{ir^{2}}{4D\tilde{\tau}}}\Gamma\left[0,\frac{ir^{2}}{4D\tilde{\tau}}\right]}{\tilde{\tau}^{2}}\right\} \\
\tilde{\tau} & =\tau+i\tau_{e}\end{align*}
and where $\Gamma\left[\alpha,z\right]$ is the incomplete gamma function.

{}

\subsection*{Expansion in terms of Legendre polynomials}

In order to investigate the scaling dimensions of the bare kernel
$\tilde{P}_{c}$ in the non-local regime, we can use the approximate
expression of Eq. (\ref{eq:Pc_approx}) \begin{align*}
\tilde{P}_{c}\left(\mathbf{x}\right) & \approx\frac{\left(4D\right)^{2}}{x^{4}+\left(4D\tau\right)^{2}}\end{align*}
 and express it in terms of cylindrical coordinates $\left\{ x=r\cos\phi,u\tau=r\sin\phi\right\} $
as

\begin{align}
\tilde{P}_{c}\left(\mathbf{x}\right) & =\tilde{P}_{c}\left(r,\phi\right)\nonumber \\
 & \approx\frac{\left(4D\right)^{2}}{r_{D}^{4}}\frac{1}{\tilde{r}^{2}}\tilde{f}\left(\tilde{r},\phi\right),\label{eq:Pc_polar}\end{align}
 where the definitions\begin{align}
\tilde{f}\left(\tilde{r},\phi\right) & \equiv\frac{1}{\tilde{r}^{2}\cos^{4}\phi+\sin^{2}\phi}\label{eq:function_f_approx}\\
\tilde{r} & \equiv\frac{r}{r_{D}}\label{eq:r_tilde}\\
r_{D} & \equiv\frac{4D}{u}=\frac{4v_{F}}{u}l_{e},\label{eq:r_D}\end{align}
has been used and where $v_{F}$ is the Fermi velocity in the metallic
film. We now expand $\tilde{f}\left(\tilde{r},\phi\right)$ in Legendre
polynomials in order to separate radial and angular variables in a
systematic way \begin{align*}
\tilde{f}\left(\tilde{r},\phi\right) & =\sum_{\ell=0}^{\infty}A_{\ell}\left(\tilde{r}\right)\mathcal{P}_{\ell}\left(\cos\phi\right)\end{align*}
 where the coefficients are defined as\begin{align}
A_{\ell}\left(\tilde{r}\right) & \equiv\frac{2\ell+1}{2}\int_{-1}^{1}d\left(\cos\phi\right)\;\tilde{f}\left(\tilde{r},\phi\right)\mathcal{P}_{\ell}\left(\cos\phi\right).\label{eq:Al_def}\end{align}
 Then,\begin{align}
\tilde{P}_{c}\left(r,\phi\right) & =\frac{\left(4D\right)^{2}}{r_{D}^{4}}\frac{1}{\tilde{r}^{2}}\sum_{\ell=0}^{\infty}A_{\ell}\left(\tilde{r}\right)\mathcal{P}_{\ell}\left(\cos\phi\right)\label{eq:Pc_auxiliar}\end{align}
 Changing variables to $\nu\equiv\cos\phi$, we can write Eq. (\ref{eq:Al_def})
as

\begin{align}
A_{\ell}\left(\tilde{r}\right) & =\frac{2\ell+1}{2}\int_{-1}^{1}d\nu\;\mathcal{P}_{\ell}\left(\nu\right)\tilde{f}\left(\tilde{r},\nu\right)\nonumber \\
 & =\frac{2\ell+1}{2}\int_{-1}^{1}d\nu\;\frac{\mathcal{P}_{\ell}\left(\nu\right)}{\tilde{r}^{2}\nu^{4}-\nu^{2}+1}.\label{eq:A_l_approx}\end{align}
An asymptotic approximation of $A_{\ell}\left(\tilde{r}\right)$ in
the regime $\tilde{r}\rightarrow\infty$ gives\begin{align}
\lim_{\tilde{r}\rightarrow\infty}A_{\ell}\left(\tilde{r}\right) & \rightarrow\frac{2\ell+1}{2}\mathcal{P}_{\ell}\left(0\right)\int_{-1}^{1}d\nu\;\frac{1}{\tilde{r}^{2}\nu^{4}+1},\nonumber \\
 & \rightarrow\frac{2\ell+1}{2}\mathcal{P}_{\ell}\left(0\right)\frac{\pi}{\sqrt{2\tilde{r}}}\label{eq:A_l_approx_2}\end{align}
where from the Rodrigues formula\citep{abramowitz} we obtain $\mathcal{P}_{\ell}\left(0\right)=\frac{\left(-1\right)^{\ell/2}}{2^{\ell}}\frac{\ell!}{\left[\left(\ell/2\right)!\right]^{2}}$
for $\ell$ even (note from Eq. (\ref{eq:A_l_approx}) that $A_{\ell}\left(\tilde{r}\right)$
vanish for odd values of $\ell$). %
\begin{comment}
This asymptotic form, together with the expansion in Eq. (\ref{eq:Pc_auxiliar})
gives a decay $\tilde{P}_{c}\left(r,\phi\right)\sim\tilde{r}^{-5/2}$
at large values of $r$.
\end{comment}
{}%
{}Therefore, from Eqs. (\ref{eq:Pc_auxiliar})and (\ref{eq:A_l_approx_2})
we obtain in the asymptotic limit %
\begin{comment}
\begin{align*}
\lim_{\tilde{r}\rightarrow\infty}A_{\ell}\left(\tilde{r}\right) & \rightarrow\frac{a_{\ell}}{\sqrt{\tilde{r}}}\quad\forall\;\ell.\end{align*}
coefficients $a_{\ell}$ are defined as\begin{align}
a_{\ell} & \equiv\frac{2\ell+1}{2}\frac{\pi}{\sqrt{2}}\mathcal{P}_{\ell}\left(0\right)\label{eq:def_a_l}\end{align}
Therefore, in the limit $\tilde{r}\rightarrow\infty$, the cooperon
propagator in Eq. (\ref{eq:Pc_auxiliar}) can be written as
\end{comment}
{}\begin{align*}
\tilde{P}_{c}\left(r,\phi\right) & \xrightarrow[r\rightarrow\infty]{}\frac{\pi}{\sqrt{2}}\frac{\left(4D\right)^{2}}{r_{D}^{4}}\frac{1}{\tilde{r}^{\frac{5}{2}}}\sum_{\ell=0}^{\infty}\frac{2\ell+1}{2}\mathcal{P}_{\ell}\left(0\right)\mathcal{P}_{\ell}\left(\cos\phi\right),\\
 & \xrightarrow[r\rightarrow\infty]{}\frac{\pi}{\sqrt{2}}\frac{\left(4D\right)^{2}}{r_{D}^{4}}\frac{1}{\tilde{r}^{\frac{5}{2}}}\delta\left(\cos\phi\right),\end{align*}
where in the last line the expansion of the Dirac-delta function in
terms of the Legendre polynomials has been used. Coming back to coordinates
$x$ and $\tau$, we obtain \begin{align*}
\tilde{P}_{c}\left(\mathbf{x}\right) & \xrightarrow[r\rightarrow\infty]{}\frac{\pi\sqrt{4D}}{\sqrt{2}}\frac{\delta\left(x\right)}{\tau^{\frac{3}{2}}},\end{align*}
which provides a rigourous derivation of Eq. (\ref{eq:Pc_approx_eff}).

\bibliographystyle{apsrev} \bibliographystyle{apsrev}

\begin{thebibliography}{52}
\expandafter\ifx\csname natexlab\endcsname\relax\def\natexlab#1{#1}\fi
\expandafter\ifx\csname bibnamefont\endcsname\relax
  \def\bibnamefont#1{#1}\fi
\expandafter\ifx\csname bibfnamefont\endcsname\relax
  \def\bibfnamefont#1{#1}\fi
\expandafter\ifx\csname citenamefont\endcsname\relax
  \def\citenamefont#1{#1}\fi
\expandafter\ifx\csname url\endcsname\relax
  \def\url#1{\texttt{#1}}\fi
\expandafter\ifx\csname urlprefix\endcsname\relax\def\urlprefix{URL }\fi
\providecommand{\bibinfo}[2]{#2}
\providecommand{\eprint}[2][]{\url{#2}}

\bibitem[{\citenamefont{Mermin and Wagner}(1966)}]{mermin_wagner_theorem}
\bibinfo{author}{\bibfnamefont{N.~D.} \bibnamefont{Mermin}} \bibnamefont{and}
  \bibinfo{author}{\bibfnamefont{H.}~\bibnamefont{Wagner}},
  \bibinfo{journal}{Phys. Rev. Lett.} \textbf{\bibinfo{volume}{17}},
  \bibinfo{pages}{1133} (\bibinfo{year}{1966}).

\bibitem[{\citenamefont{Giamarchi}(2004)}]{giamarchi_book_1d}
\bibinfo{author}{\bibfnamefont{T.}~\bibnamefont{Giamarchi}},
  \emph{\bibinfo{title}{Quantum Physics in One Dimension}}
  (\bibinfo{publisher}{Oxford University Press}, \bibinfo{address}{Oxford},
  \bibinfo{year}{2004}).

\bibitem[{\citenamefont{Caldeira and Leggett}(1983)}]{caldeira_leggett}
\bibinfo{author}{\bibfnamefont{A.~O.} \bibnamefont{Caldeira}} \bibnamefont{and}
  \bibinfo{author}{\bibfnamefont{A.~J.} \bibnamefont{Leggett}},
  \bibinfo{journal}{Ann. Phys. (N. Y.)} \textbf{\bibinfo{volume}{149}},
  \bibinfo{pages}{374} (\bibinfo{year}{1983}).

\bibitem[{\citenamefont{Bray and
  Moore}(1982)}]{bray82_macroscopic_quantum_tunneling}
\bibinfo{author}{\bibfnamefont{A.~J.} \bibnamefont{Bray}} \bibnamefont{and}
  \bibinfo{author}{\bibfnamefont{M.~A.} \bibnamefont{Moore}},
  \bibinfo{journal}{Phys. Rev. Lett.} \textbf{\bibinfo{volume}{49}},
  \bibinfo{pages}{1545} (\bibinfo{year}{1982}).

\bibitem[{\citenamefont{Chakravarty}(1982)}]{chakravarty82_macroscopic_quantum%
_tunneling}
\bibinfo{author}{\bibfnamefont{S.}~\bibnamefont{Chakravarty}},
  \bibinfo{journal}{Phys. Rev. Lett.} \textbf{\bibinfo{volume}{49}},
  \bibinfo{pages}{681} (\bibinfo{year}{1982}).

\bibitem[{\citenamefont{Schmid}(1983)}]{schmid83_diffusion_dissipative_quantum%
_system}
\bibinfo{author}{\bibfnamefont{A.}~\bibnamefont{Schmid}},
  \bibinfo{journal}{Phys. Rev. Lett.} \textbf{\bibinfo{volume}{51}},
  \bibinfo{pages}{1506} (\bibinfo{year}{1983}).

\bibitem[{\citenamefont{Sch{\"o}n and
  Zaikin}(1990)}]{schoen90_review_ultrasmall_tunnel_junctions}
\bibinfo{author}{\bibfnamefont{G.}~\bibnamefont{Sch{\"o}n}} \bibnamefont{and}
  \bibinfo{author}{\bibfnamefont{A.~D.} \bibnamefont{Zaikin}},
  \bibinfo{journal}{Physics Reports} \textbf{\bibinfo{volume}{198}},
  \bibinfo{pages}{237 } (\bibinfo{year}{1990}), ISSN \bibinfo{issn}{0370-1573},
  \urlprefix\url{http://www.sciencedirect.com/science/article/B6TVP-46T4WMC-6V%
/2/56e621d5b6c3e08daef3b06fb8b5d3c6}.

\bibitem[{\citenamefont{Fazio and van~der
  Zant}(2001)}]{fazio01_review_superconducting_networks}
\bibinfo{author}{\bibfnamefont{R.}~\bibnamefont{Fazio}} \bibnamefont{and}
  \bibinfo{author}{\bibfnamefont{H.}~\bibnamefont{van~der Zant}},
  \bibinfo{journal}{Physics Reports} \textbf{\bibinfo{volume}{355}},
  \bibinfo{pages}{235} (\bibinfo{year}{2001}).

\bibitem[{\citenamefont{Bobbert et~al.}(1992)\citenamefont{Bobbert, Fazio,
  Sch{\"o}n, and Zaikin}}]{bobbert92_transitions_dissipative_josephson_chains}
\bibinfo{author}{\bibfnamefont{P.~A.} \bibnamefont{Bobbert}},
  \bibinfo{author}{\bibfnamefont{R.}~\bibnamefont{Fazio}},
  \bibinfo{author}{\bibfnamefont{G.}~\bibnamefont{Sch{\"o}n}},
  \bibnamefont{and} \bibinfo{author}{\bibfnamefont{A.~D.}
  \bibnamefont{Zaikin}}, \bibinfo{journal}{Phys. Rev. B}
  \textbf{\bibinfo{volume}{45}}, \bibinfo{pages}{2294} (\bibinfo{year}{1992}).

\bibitem[{\citenamefont{Goswami and
  Chakravarty}(2006)}]{goswami06_josephson_array}
\bibinfo{author}{\bibfnamefont{P.}~\bibnamefont{Goswami}} \bibnamefont{and}
  \bibinfo{author}{\bibfnamefont{S.}~\bibnamefont{Chakravarty}},
  \bibinfo{journal}{Phys. Rev. B} \textbf{\bibinfo{volume}{73}},
  \bibinfo{pages}{094516} (\bibinfo{year}{2006}).

\bibitem[{\citenamefont{Refael et~al.}(2007)\citenamefont{Refael, Demler, Oreg,
  and Fisher}}]{refael07_SN_transition_in_grains&nanowires}
\bibinfo{author}{\bibfnamefont{G.}~\bibnamefont{Refael}},
  \bibinfo{author}{\bibfnamefont{E.}~\bibnamefont{Demler}},
  \bibinfo{author}{\bibfnamefont{Y.}~\bibnamefont{Oreg}}, \bibnamefont{and}
  \bibinfo{author}{\bibfnamefont{D.~S.} \bibnamefont{Fisher}},
  \bibinfo{journal}{Phys. Rev. B} \textbf{\bibinfo{volume}{75}},
  \bibinfo{pages}{014522} (\bibinfo{year}{2007}).

\bibitem[{\citenamefont{Feigel'man and
  Larkin}(1998)}]{feigelman98_smt_in_2d_proximity_array}
\bibinfo{author}{\bibfnamefont{M.~V.} \bibnamefont{Feigel'man}}
  \bibnamefont{and} \bibinfo{author}{\bibfnamefont{A.~I.}
  \bibnamefont{Larkin}}, \bibinfo{journal}{Chem. Phys.}
  \textbf{\bibinfo{volume}{235}}, \bibinfo{pages}{107} (\bibinfo{year}{1998}).

\bibitem[{\citenamefont{Feigel'man et~al.}(2000)\citenamefont{Feigel'man,
  Larkin, and Skvortsov}}]{feigelman00_keldysh_action_superconductor}
\bibinfo{author}{\bibfnamefont{M.~V.} \bibnamefont{Feigel'man}},
  \bibinfo{author}{\bibfnamefont{A.~I.} \bibnamefont{Larkin}},
  \bibnamefont{and} \bibinfo{author}{\bibfnamefont{M.~A.}
  \bibnamefont{Skvortsov}}, \bibinfo{journal}{Phys. Rev. B}
  \textbf{\bibinfo{volume}{61}}, \bibinfo{pages}{12361} (\bibinfo{year}{2000}).

\bibitem[{\citenamefont{Spivak et~al.}(2001)\citenamefont{Spivak, Zyuzin, and
  Hruska}}]{spivak01_quantum_superconductor_metal_transition}
\bibinfo{author}{\bibfnamefont{B.}~\bibnamefont{Spivak}},
  \bibinfo{author}{\bibfnamefont{A.}~\bibnamefont{Zyuzin}}, \bibnamefont{and}
  \bibinfo{author}{\bibfnamefont{M.}~\bibnamefont{Hruska}},
  \bibinfo{journal}{Phys. Rev. B} \textbf{\bibinfo{volume}{64}},
  \bibinfo{pages}{132502} (\bibinfo{year}{2001}).

\bibitem[{\citenamefont{Feigel'man et~al.}(2001)\citenamefont{Feigel'man,
  Larkin, and Skvortsov}}]{feigelman01_smt_proximity_array}
\bibinfo{author}{\bibfnamefont{M.~V.} \bibnamefont{Feigel'man}},
  \bibinfo{author}{\bibfnamefont{A.~I.} \bibnamefont{Larkin}},
  \bibnamefont{and} \bibinfo{author}{\bibfnamefont{M.~A.}
  \bibnamefont{Skvortsov}}, \bibinfo{journal}{Phys. Rev. Lett.}
  \textbf{\bibinfo{volume}{86}}, \bibinfo{pages}{1869} (\bibinfo{year}{2001}).

\bibitem[{\citenamefont{{Castro Neto} et~al.}(1997)\citenamefont{{Castro Neto},
  de~C.~Chamon, and Nayak}}]{castroneto97_open_luttinger_liquids}
\bibinfo{author}{\bibfnamefont{A.~H.} \bibnamefont{{Castro Neto}}},
  \bibinfo{author}{\bibfnamefont{C.}~\bibnamefont{de~C.~Chamon}},
  \bibnamefont{and} \bibinfo{author}{\bibfnamefont{C.}~\bibnamefont{Nayak}},
  \bibinfo{journal}{Phys. Rev. Lett.} \textbf{\bibinfo{volume}{79}},
  \bibinfo{pages}{4629} (\bibinfo{year}{1997}).

\bibitem[{\citenamefont{Cazalilla et~al.}(2006)\citenamefont{Cazalilla, Sols,
  and Guinea}}]{cazalilla06_dissipative_transition}
\bibinfo{author}{\bibfnamefont{M.~A.} \bibnamefont{Cazalilla}},
  \bibinfo{author}{\bibfnamefont{F.}~\bibnamefont{Sols}}, \bibnamefont{and}
  \bibinfo{author}{\bibfnamefont{F.}~\bibnamefont{Guinea}},
  \bibinfo{journal}{Phys. Rev. Lett.} \textbf{\bibinfo{volume}{97}},
  \bibinfo{pages}{076401} (\bibinfo{year}{2006}).

\bibitem[{\citenamefont{Artemenko and
  Nattermann}(2007)}]{artemenko07_longrange_order_in_1d}
\bibinfo{author}{\bibfnamefont{S.~N.} \bibnamefont{Artemenko}}
  \bibnamefont{and}
  \bibinfo{author}{\bibfnamefont{T.}~\bibnamefont{Nattermann}},
  \bibinfo{journal}{Phys. Rev. Lett.} \textbf{\bibinfo{volume}{99}},
  \bibinfo{pages}{256401} (\bibinfo{year}{2007}).

\bibitem[{\citenamefont{Little}(1967)}]{little67_phase_slip}
\bibinfo{author}{\bibfnamefont{W.}~\bibnamefont{Little}},
  \bibinfo{journal}{Phys. Rev.} \textbf{\bibinfo{volume}{156}},
  \bibinfo{pages}{396} (\bibinfo{year}{1967}).

\bibitem[{\citenamefont{Langer and Ambegaokar}(1967)}]{langer67}
\bibinfo{author}{\bibfnamefont{J.~S.} \bibnamefont{Langer}} \bibnamefont{and}
  \bibinfo{author}{\bibfnamefont{V.}~\bibnamefont{Ambegaokar}},
  \bibinfo{journal}{Phys. Rev.} \textbf{\bibinfo{volume}{164}},
  \bibinfo{pages}{498} (\bibinfo{year}{1967}).

\bibitem[{\citenamefont{McCumber and Halperin}(1970)}]{mccumber70}
\bibinfo{author}{\bibfnamefont{D.~E.} \bibnamefont{McCumber}} \bibnamefont{and}
  \bibinfo{author}{\bibfnamefont{B.~I.} \bibnamefont{Halperin}},
  \bibinfo{journal}{Phys. Rev. B} \textbf{\bibinfo{volume}{1}},
  \bibinfo{pages}{1054} (\bibinfo{year}{1970}).

\bibitem[{\citenamefont{Giordano}(1988)}]{giordano88_evidence_of_qps}
\bibinfo{author}{\bibfnamefont{N.}~\bibnamefont{Giordano}},
  \bibinfo{journal}{Phys. Rev. Lett.} \textbf{\bibinfo{volume}{61}},
  \bibinfo{pages}{2137} (\bibinfo{year}{1988}).

\bibitem[{\citenamefont{Giordano}(1994)}]{giordano94}
\bibinfo{author}{\bibfnamefont{N.}~\bibnamefont{Giordano}},
  \bibinfo{journal}{Physica B} \textbf{\bibinfo{volume}{203}},
  \bibinfo{pages}{460} (\bibinfo{year}{1994}).

\bibitem[{\citenamefont{Lau et~al.}(2001)\citenamefont{Lau, Markovic, Bockrath,
  Bezryadin, and Tinkham}}]{lau01}
\bibinfo{author}{\bibfnamefont{C.~N.} \bibnamefont{Lau}},
  \bibinfo{author}{\bibfnamefont{N.}~\bibnamefont{Markovic}},
  \bibinfo{author}{\bibfnamefont{M.}~\bibnamefont{Bockrath}},
  \bibinfo{author}{\bibfnamefont{A.}~\bibnamefont{Bezryadin}},
  \bibnamefont{and} \bibinfo{author}{\bibfnamefont{M.}~\bibnamefont{Tinkham}},
  \bibinfo{journal}{Phys. Rev. Lett.} \textbf{\bibinfo{volume}{87}},
  \bibinfo{pages}{217003} (\bibinfo{year}{2001}).

\bibitem[{\citenamefont{Tian et~al.}(2005)\citenamefont{Tian, Wang, Kurtz, Liu,
  Chan, Mayer, and Mallouk}}]{tian05_dissipation_in_Sn_wires}
\bibinfo{author}{\bibfnamefont{M.}~\bibnamefont{Tian}},
  \bibinfo{author}{\bibfnamefont{J.}~\bibnamefont{Wang}},
  \bibinfo{author}{\bibfnamefont{J.~S.} \bibnamefont{Kurtz}},
  \bibinfo{author}{\bibfnamefont{Y.}~\bibnamefont{Liu}},
  \bibinfo{author}{\bibfnamefont{M.~H.} \bibnamefont{Chan}},
  \bibinfo{author}{\bibfnamefont{T.~S.} \bibnamefont{Mayer}}, \bibnamefont{and}
  \bibinfo{author}{\bibfnamefont{T.~E.} \bibnamefont{Mallouk}},
  \bibinfo{journal}{Phys. Rev. B} \textbf{\bibinfo{volume}{71}},
  \bibinfo{pages}{104521} (\bibinfo{year}{2005}).

\bibitem[{\citenamefont{Zgirski et~al.}(2005)\citenamefont{Zgirski, Riikonen,
  Touboltsev, and
  Arutyunov}}]{zgirski05_breakdown_of_superconductivity_in_wires}
\bibinfo{author}{\bibfnamefont{M.}~\bibnamefont{Zgirski}},
  \bibinfo{author}{\bibfnamefont{K.-P.} \bibnamefont{Riikonen}},
  \bibinfo{author}{\bibfnamefont{V.}~\bibnamefont{Touboltsev}},
  \bibnamefont{and}
  \bibinfo{author}{\bibfnamefont{K.}~\bibnamefont{Arutyunov}},
  \bibinfo{journal}{Nano Lett.} \textbf{\bibinfo{volume}{5}},
  \bibinfo{pages}{1029} (\bibinfo{year}{2005}).

\bibitem[{\citenamefont{Altomare et~al.}(2006)\citenamefont{Altomare, Chang,
  Melloch, Hong, and Tu}}]{altomare06_experimental_evidence_of_qps}
\bibinfo{author}{\bibfnamefont{F.}~\bibnamefont{Altomare}},
  \bibinfo{author}{\bibfnamefont{A.~M.} \bibnamefont{Chang}},
  \bibinfo{author}{\bibfnamefont{M.~R.} \bibnamefont{Melloch}},
  \bibinfo{author}{\bibfnamefont{Y.}~\bibnamefont{Hong}}, \bibnamefont{and}
  \bibinfo{author}{\bibfnamefont{C.~W.} \bibnamefont{Tu}},
  \bibinfo{journal}{Phys. Rev. Lett.} \textbf{\bibinfo{volume}{97}},
  \bibinfo{pages}{017001} (\bibinfo{year}{2006}).

\bibitem[{\citenamefont{Bollinger et~al.}(2008)\citenamefont{Bollinger,
  {Dinsmore III}, Rogachev, and Bezryadin}}]{bollinger08_sit_diagram_for_wires}
\bibinfo{author}{\bibfnamefont{A.~T.} \bibnamefont{Bollinger}},
  \bibinfo{author}{\bibfnamefont{R.~C.} \bibnamefont{{Dinsmore III}}},
  \bibinfo{author}{\bibfnamefont{A.}~\bibnamefont{Rogachev}}, \bibnamefont{and}
  \bibinfo{author}{\bibfnamefont{A.}~\bibnamefont{Bezryadin}},
  \bibinfo{journal}{Phys. Rev. Lett.} \textbf{\bibinfo{volume}{101}},
  \bibinfo{pages}{227003} (\bibinfo{year}{2008}).

\bibitem[{\citenamefont{Giamarchi}(1992)}]{giamarchi_attract_1d}
\bibinfo{author}{\bibfnamefont{T.}~\bibnamefont{Giamarchi}},
  \bibinfo{journal}{Phys. Rev. B} \textbf{\bibinfo{volume}{46}},
  \bibinfo{pages}{342} (\bibinfo{year}{1992}).

\bibitem[{\citenamefont{Zaikin et~al.}(1997)\citenamefont{Zaikin, Golubev, van
  Otterlo, and Zim{\'a}nyi}}]{zaikin97}
\bibinfo{author}{\bibfnamefont{A.~D.} \bibnamefont{Zaikin}},
  \bibinfo{author}{\bibfnamefont{D.~S.} \bibnamefont{Golubev}},
  \bibinfo{author}{\bibfnamefont{A.}~\bibnamefont{van Otterlo}},
  \bibnamefont{and} \bibinfo{author}{\bibfnamefont{G.~T.}
  \bibnamefont{Zim{\'a}nyi}}, \bibinfo{journal}{Phys. Rev. Lett.}
  \textbf{\bibinfo{volume}{78}}, \bibinfo{pages}{1552} (\bibinfo{year}{1997}).

\bibitem[{\citenamefont{Bezryadin et~al.}(2000)\citenamefont{Bezryadin, Lau,
  and Tinkham}}]{bezryadin00}
\bibinfo{author}{\bibfnamefont{A.}~\bibnamefont{Bezryadin}},
  \bibinfo{author}{\bibfnamefont{C.~N.} \bibnamefont{Lau}}, \bibnamefont{and}
  \bibinfo{author}{\bibfnamefont{M.}~\bibnamefont{Tinkham}},
  \bibinfo{journal}{Nature} \textbf{\bibinfo{volume}{404}},
  \bibinfo{pages}{971} (\bibinfo{year}{2000}).

\bibitem[{\citenamefont{B{\"u}chler et~al.}(2004)\citenamefont{B{\"u}chler,
  Geshkenbein, and Blatter}}]{buchler04_sit_finite_length_wire}
\bibinfo{author}{\bibfnamefont{H.~P.} \bibnamefont{B{\"u}chler}},
  \bibinfo{author}{\bibfnamefont{V.~B.} \bibnamefont{Geshkenbein}},
  \bibnamefont{and} \bibinfo{author}{\bibfnamefont{G.}~\bibnamefont{Blatter}},
  \bibinfo{journal}{Phys. Rev. Lett.} \textbf{\bibinfo{volume}{92}},
  \bibinfo{pages}{067007} (\bibinfo{year}{2004}).

\bibitem[{\citenamefont{Arutyunov et~al.}(2008)\citenamefont{Arutyunov,
  Golubev, and Zaikin}}]{arutyunov08_superconductivity_1d_review}
\bibinfo{author}{\bibfnamefont{K.~Y.} \bibnamefont{Arutyunov}},
  \bibinfo{author}{\bibfnamefont{D.~S.} \bibnamefont{Golubev}},
  \bibnamefont{and} \bibinfo{author}{\bibfnamefont{A.~D.}
  \bibnamefont{Zaikin}}, \bibinfo{journal}{Phys. Rep.}
  \textbf{\bibinfo{volume}{464}}, \bibinfo{pages}{1} (\bibinfo{year}{2008}).

\bibitem[{\citenamefont{Kosterlitz and Thouless}(1973)}]{kosterlitz_thouless}
\bibinfo{author}{\bibfnamefont{J.~M.} \bibnamefont{Kosterlitz}}
  \bibnamefont{and} \bibinfo{author}{\bibfnamefont{D.~J.}
  \bibnamefont{Thouless}}, \bibinfo{journal}{J. Phys. C}
  \textbf{\bibinfo{volume}{6}}, \bibinfo{pages}{1181} (\bibinfo{year}{1973}).

\bibitem[{\citenamefont{Hekking and Nazarov}(1993)}]{hekking93}
\bibinfo{author}{\bibfnamefont{F.~W.} \bibnamefont{Hekking}} \bibnamefont{and}
  \bibinfo{author}{\bibfnamefont{Y.~V.} \bibnamefont{Nazarov}},
  \bibinfo{journal}{Phys. Rev. Lett.} \textbf{\bibinfo{volume}{71}},
  \bibinfo{pages}{1625} (\bibinfo{year}{1993}).

\bibitem[{\citenamefont{Hekking and
  Nazarov}(1994)}]{hekking94_andreev_conductance}
\bibinfo{author}{\bibfnamefont{F.~W.} \bibnamefont{Hekking}} \bibnamefont{and}
  \bibinfo{author}{\bibfnamefont{Y.~V.} \bibnamefont{Nazarov}},
  \bibinfo{journal}{Phys. Rev. B} \textbf{\bibinfo{volume}{49}},
  \bibinfo{pages}{6847} (\bibinfo{year}{1994}).

\bibitem[{\citenamefont{Aitchison et~al.}(1995)\citenamefont{Aitchison, Ao,
  Thouless, and Zhu}}]{aitchison95_effective_bcs_lagragian}
\bibinfo{author}{\bibfnamefont{I.~J.~R.} \bibnamefont{Aitchison}},
  \bibinfo{author}{\bibfnamefont{P.}~\bibnamefont{Ao}},
  \bibinfo{author}{\bibfnamefont{D.~J.} \bibnamefont{Thouless}},
  \bibnamefont{and} \bibinfo{author}{\bibfnamefont{X.-M.} \bibnamefont{Zhu}},
  \bibinfo{journal}{Phys. Rev. B} \textbf{\bibinfo{volume}{51}},
  \bibinfo{pages}{6531} (\bibinfo{year}{1995}).

\bibitem[{\citenamefont{van Otterlo et~al.}(1999)\citenamefont{van Otterlo,
  Golubev, Zaikin, and Blatter}}]{vanotterlo98}
\bibinfo{author}{\bibfnamefont{A.}~\bibnamefont{van Otterlo}},
  \bibinfo{author}{\bibfnamefont{D.~S.} \bibnamefont{Golubev}},
  \bibinfo{author}{\bibfnamefont{A.}~\bibnamefont{Zaikin}}, \bibnamefont{and}
  \bibinfo{author}{\bibfnamefont{G.}~\bibnamefont{Blatter}},
  \bibinfo{journal}{Eur. Phys. J. B} \textbf{\bibinfo{volume}{10}},
  \bibinfo{pages}{131} (\bibinfo{year}{1999}).

\bibitem[{\citenamefont{Fetter and Walecka}(1971)}]{fetter}
\bibinfo{author}{\bibfnamefont{A.~L.} \bibnamefont{Fetter}} \bibnamefont{and}
  \bibinfo{author}{\bibfnamefont{J.~D.} \bibnamefont{Walecka}},
  \emph{\bibinfo{title}{Quantum theory of many-particle systems}}
  (\bibinfo{publisher}{McGraw-Hill}, \bibinfo{address}{New York},
  \bibinfo{year}{1971}).

\bibitem[{\citenamefont{Efetov and
  Larkin}(1974)}]{efetov74_fluctuations_1d_superconductor}
\bibinfo{author}{\bibfnamefont{K.~B.} \bibnamefont{Efetov}} \bibnamefont{and}
  \bibinfo{author}{\bibfnamefont{A.~I.} \bibnamefont{Larkin}},
  \bibinfo{journal}{Sov. Phys. JETP} \textbf{\bibinfo{volume}{39}},
  \bibinfo{pages}{1129} (\bibinfo{year}{1974}).

\bibitem[{\citenamefont{Mooij and Sch{\"o}n}(1985)}]{mooij85_mooij_schon_mode}
\bibinfo{author}{\bibfnamefont{J.~E.} \bibnamefont{Mooij}} \bibnamefont{and}
  \bibinfo{author}{\bibfnamefont{G.}~\bibnamefont{Sch{\"o}n}},
  \bibinfo{journal}{Phys. Rev. Lett.} \textbf{\bibinfo{volume}{55}},
  \bibinfo{pages}{114} (\bibinfo{year}{1985}).

\bibitem[{\citenamefont{Bruder et~al.}(1994)\citenamefont{Bruder, Fazio, and
  Sch{\"o}n}}]{bruder94}
\bibinfo{author}{\bibfnamefont{C.}~\bibnamefont{Bruder}},
  \bibinfo{author}{\bibfnamefont{R.}~\bibnamefont{Fazio}}, \bibnamefont{and}
  \bibinfo{author}{\bibfnamefont{G.}~\bibnamefont{Sch{\"o}n}},
  \bibinfo{journal}{Physica B} \textbf{\bibinfo{volume}{203}},
  \bibinfo{pages}{240} (\bibinfo{year}{1994}).

\bibitem[{\citenamefont{Akkermans and Montambaux}(2004)}]{akkermans}
\bibinfo{author}{\bibfnamefont{E.}~\bibnamefont{Akkermans}} \bibnamefont{and}
  \bibinfo{author}{\bibfnamefont{G.}~\bibnamefont{Montambaux}},
  \emph{\bibinfo{title}{Physique m\'esoscopique des \'electrons et des
  photons}} (\bibinfo{publisher}{EDP Sciences/CNRS}, \bibinfo{address}{Paris},
  \bibinfo{year}{2004}).

\bibitem[{\citenamefont{Tinkham}(1996)}]{tinkham}
\bibinfo{author}{\bibfnamefont{M.}~\bibnamefont{Tinkham}},
  \emph{\bibinfo{title}{Introduction to Superconductivity}}
  (\bibinfo{publisher}{McGraw-Hill}, \bibinfo{year}{1996}),
  \bibinfo{edition}{2nd} ed.

\bibitem[{lob()}]{lobos09}
\bibinfo{note}{A. M. Lobos, A. Iucci, M. M{\"u}ller, and T. Giamarchi, (in
  preparation)}.

\bibitem[{\citenamefont{{De Gennes}}(1966)}]{DeGennes_supra}
\bibinfo{author}{\bibfnamefont{P.~G.} \bibnamefont{{De Gennes}}}, \emph{\bibinfo{title}{Superconductivity of Metals and Alloys}}
  (\bibinfo{publisher}{W. A. Benjamin}, \bibinfo{address}{New York},
  \bibinfo{year}{1966}).

\bibitem[{\citenamefont{Feynman}(1972)}]{feynman_statmech}
\bibinfo{author}{\bibfnamefont{R.~P.} \bibnamefont{Feynman}},
  \emph{\bibinfo{title}{Statistical Mechanics}} (\bibinfo{publisher}{Benjamin},
  \bibinfo{address}{Reading, MA}, \bibinfo{year}{1972}).

\bibitem[{\citenamefont{G{\"o}tze and
  W{\"o}lfle}(1972)}]{gotze_fonction_memoire}
\bibinfo{author}{\bibfnamefont{W.}~\bibnamefont{G{\"o}tze}} \bibnamefont{and}
  \bibinfo{author}{\bibfnamefont{P.}~\bibnamefont{W{\"o}lfle}},
  \bibinfo{journal}{Phys. Rev. B} \textbf{\bibinfo{volume}{6}},
  \bibinfo{pages}{1226} (\bibinfo{year}{1972}).

\bibitem[{\citenamefont{Merchant et~al.}(2001)\citenamefont{Merchant, Ostrick,
  Barber, and Dynes}}]{merchant01_crossover_phase_amplitude_fluctuations}
\bibinfo{author}{\bibfnamefont{L.}~\bibnamefont{Merchant}},
  \bibinfo{author}{\bibfnamefont{J.}~\bibnamefont{Ostrick}},
  \bibinfo{author}{\bibfnamefont{R.~P.} \bibnamefont{Barber}},
  \bibnamefont{and} \bibinfo{author}{\bibfnamefont{R.~C.} \bibnamefont{Dynes}},
  \bibinfo{journal}{Phys. Rev. B} \textbf{\bibinfo{volume}{63}},
  \bibinfo{pages}{134508} (\bibinfo{year}{2001}).

\bibitem[{\citenamefont{Berg et~al.}(2008)\citenamefont{Berg, Orgad, and
  Kivelson}}]{berg08_route_to_htcs_composite_systems}
\bibinfo{author}{\bibfnamefont{E.}~\bibnamefont{Berg}},
  \bibinfo{author}{\bibfnamefont{D.}~\bibnamefont{Orgad}}, \bibnamefont{and}
  \bibinfo{author}{\bibfnamefont{S.~A.} \bibnamefont{Kivelson}},
  \bibinfo{journal}{Phys. Rev. B} \textbf{\bibinfo{volume}{78}},
  \bibinfo{pages}{094509} (\bibinfo{year}{2008}).

\bibitem[{mor()}]{morpurgo09_private_communication}
\bibinfo{note}{A. Morpurgo, private communication}.

\bibitem[{\citenamefont{Abramowitz and Stegun}(1965)}]{abramowitz}
\bibinfo{author}{\bibfnamefont{M.}~\bibnamefont{Abramowitz}} \bibnamefont{and}
  \bibinfo{author}{\bibfnamefont{I.~A.} \bibnamefont{Stegun}},
  \emph{\bibinfo{title}{Handbook of mathematical functions : with formulas,
  graphs and mathematical tables}} (\bibinfo{publisher}{Dover},
  \bibinfo{address}{New York}, \bibinfo{year}{1965}).

\end{thebibliography}

\end{document}